\documentclass[twocolumn,prx,floatfix,superscriptaddress]{revtex4-1}
\usepackage{amsfonts,amsmath,amssymb,color,times,graphicx}
\definecolor{darkblue}{rgb}{0, 0, 0.8}
\usepackage[colorlinks=true, breaklinks=true, linkcolor=blue, citecolor=darkblue, urlcolor=darkblue]{hyperref}
\usepackage{graphicx}
\usepackage{bmpsize}
\usepackage{amsfonts}
\usepackage{gensymb}
\usepackage{bm}
\usepackage{braket}
\usepackage{mathtools}

\newcommand{\ccs}{CsCeSe$_2$}

\begin{document}
\title{Quantum Spin Dynamics Due to Strong Kitaev Interactions in the Triangular-Lattice Antiferromagnet~\ccs}
\author{Tao Xie}
\thanks{Corresponding author: xiet69@mail.sysu.edu.cn}
\thanks{These authors contributed equally to this work}
\affiliation{Center for Neutron Science and Technology, Guangdong Provincial Key Laboratory of Magnetoelectric Physics and Devices, School of Physics, Sun Yat-sen University, Guangzhou, Guangdong 510275, China}
\affiliation{Neutron Scattering Division, Oak Ridge National Laboratory, Oak Ridge, Tennessee 37831, USA}
\author{S. Gozel}
\thanks{These authors contributed equally to this work}
\affiliation{Laboratory for Theoretical and Computational Physics, Paul Scherrer Institute, CH-5232 Villigen-PSI, Switzerland}
\author{Jie Xing}
\affiliation{Materials Science and Technology Division, Oak Ridge National Laboratory, Oak Ridge, Tennessee 37831, USA}
\author{N. Zhao}
\affiliation{Department of Physics, Southern University of Science and Technology, Shenzhen, Guangdong, 518055, China.}
\author{S. M. Avdoshenko}
\affiliation{Leibniz-Institut f\"ur Festk\"orper- und Werkstoffforschung (IFW Dresden), Helmholtzstra{\ss}e 20, 01069
Dresden, Germany}
\author{L. Wu}
\affiliation{Department of Physics \&\ Academy for Advanced Interdisciplinary Studies, Southern University of Science \&\ Technology, Shenzhen, Guangdong, 518055, China}
\author{Athena~S.~Sefat}
\affiliation{Materials Science and Technology Division, Oak Ridge National Laboratory, Oak Ridge, Tennessee 37831, USA}
\author{A. L. Chernyshev}
\affiliation{Department of Physics and Astronomy, University of California, Irvine, California 92697, USA}
\author{A. M. L\"auchli}
\affiliation{Laboratory for Theoretical and Computational Physics, Paul Scherrer Institute, CH-5232 Villigen-PSI, Switzerland}
\affiliation{Institute of Physics, Ecole Polytechnique F\'ed\'erale de Lausanne (EPFL), CH-1015 Lausanne, Switzerland}
\author{A.~Podlesnyak}
\affiliation{Neutron Scattering Division, Oak Ridge National Laboratory, Oak Ridge, Tennessee 37831, USA}
\author{S.~E.~Nikitin}
\thanks{Corresponding author: stanislav.nikitin@psi.ch}
\affiliation{Laboratory for Neutron Scattering and Imaging, Paul Scherrer Institut, CH-5232 Villigen-PSI, Switzerland}

\begin{abstract}
The extraordinary properties of the Kitaev model have motivated an intense search for new physics in materials that combine geometrical and bond frustration. In this work, we employ inelastic neutron scattering, spin wave theory, and exact diagonalization to study the spin dynamics in the perfect triangular-lattice antiferromagnet (TLAF) \ccs. This material orders into a stripe phase, which is demonstrated to arise as a consequence of the off-diagonal bond-dependent terms in the spin Hamiltonian. By studying the spin dynamics at intermediate fields, we identify an interaction between the single-magnon state and the two-magnon continuum that causes decay of coherent magnon excitations, level repulsion, and transfer of spectral weight to the continuum that are controlled by the strength of the magnetic field. Our results provide a microscopic mechanism for the stabilization of the stripe phase in TLAF and show how complex many-body physics can be present in the spin dynamics in a magnet with strong Kitaev coupling even in an ordered ground state.
\end{abstract}

\maketitle

\textit{Introduction}.--- In the quantum theory of solids, elementary excitations can be described as emergent entities known as quasiparticles~\cite{prabhakar2009spin,SINHA1988xi}. Noninteracting quasiparticles have infinite lifetime and well-defined dispersion in the momentum-energy space. Interaction between quasiparticles can reduce their lifetime via decays, renormalize their dispersion, and even drive phase transitions~\cite{li2015orbitally, hong2019phase}.
Magnetic insulating systems provide ideal platforms for investigating interactions between the quasiparticles, because of their relative simplicity and purity~\cite{inosov2018quantum}. Magnetic Hamiltonians in such systems are usually dominated by several nearest-neighbor (NN) couplings~\cite{stevens1976exchange}, and can often be determined with good fidelity~\cite{vasiliev2018milestones}. Moreover, the magnetic field provides a clear nonthermal control parameter to tune the ground state of the system~\cite{giamarchi2008bose}.

\begin{figure*}[tb]
\center{\includegraphics[width=.95\linewidth]{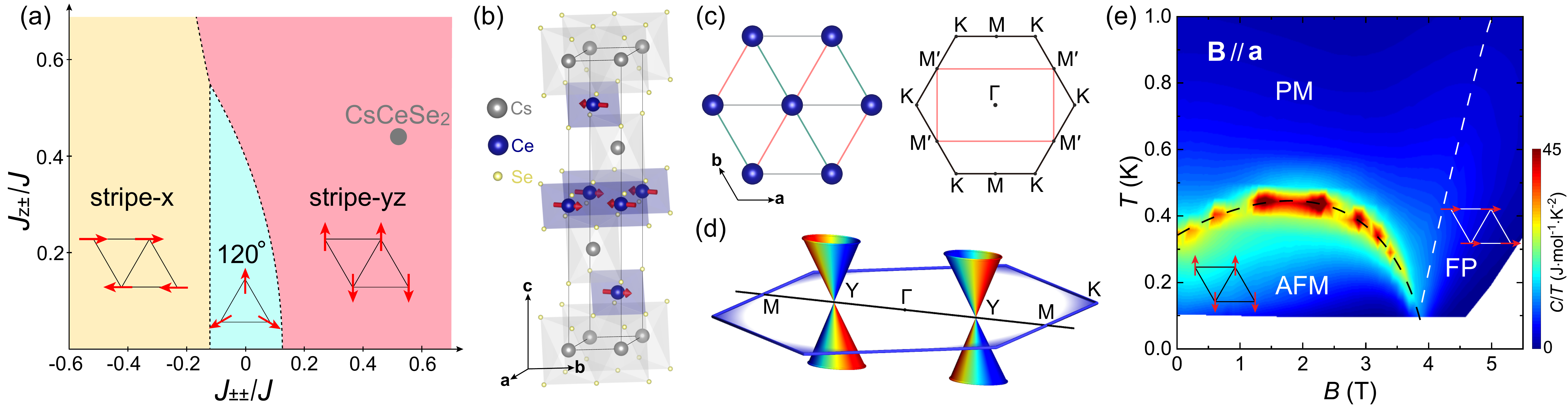}}
  \caption{(a)~Zero-temperature phase diagram of the NN $S$ = 1/2 TLAF~\eqref{Hamiltonian} with $\Delta$ = 1/4~\cite{zhu2018topography}. In the stripe-$x$ phase, spins lie in the $ab$ plane, parallel to the bonds, while in the stripe-$yz$ phase, they point perpendicular to the bonds and can be canted out of the $ab$ plane.
  (b)~Crystal and magnetic structure of \ccs. The red arrows represent magnetic moments of Ce$^{3+}$.
  (c)~Left: projection of the lattice on the $ab$ plane with three different bonds marked by different colors. Right: crystallographic (black) and magnetic (red) BZ of \ccs.
  (d)~Schematic representation of Dirac cones at the Y points, where magnons cross due to the symmetry of the stripe-$yz$ state.
  (e)~Magnetic $B$-$T$ phase diagram~\cite{Xie2023stripe}. The black dashed line represents the phase boundary between the AFM and paramagnetic (PM)/FP phase. The white dashed line indicates a crossover between the PM and FP phases.
  }
  \label{Fig1}
  \vspace{-12pt}
\end{figure*}

The magnon-magnon interaction can be restricted by the high symmetry of the system~\cite{zhitomirsky2013colloquium}, making materials with anisotropic-exchange interactions especially appealing. Because of the lattice symmetry, the NN exchange matrix of a triangular-lattice antiferromagnet (TLAF) is parameterized by four parameters: isotropic exchange $J$, $XXZ$ anisotropy $\Delta$, and two bond-dependent (BD) terms, $J_{\pm\pm}$ and $J_{z\pm}$~\cite{li2015rare,maksimov2019anisotropic}, as we will detail below. Notably, a linear combination of $J_{\pm\pm}$ and $J_{z\pm}$ constitutes the best known BD interactions, --- the celebrated Kitaev term, $KS_i^{\alpha}S_j^{\alpha}$ ($\alpha$$\in$$\{x,y,z\}$)~\cite{maksimov2022Erratum,Kitaev2006,liu2018selective}. The phase diagram of the BD TLAF has been characterized by analytical and numerical methods~\cite{Liyaodong2016,Luo2017ground,zhu2018topography,maksimov2019anisotropic,Wu2021} and is summarized in Fig.~\ref{Fig1}(a) for $\Delta$ = 1/4. It can be divided into regions of the stripe-$x$, stripe-$yz$, and 120$\degree$ order, depending on the sign of $J_{\pm\pm}$ and the value of $J_{z\pm}$. In addition, a $\mathrm{U}(1)$ quantum spin-liquid (QSL) phase has been suggested between the 120$\degree$ and stripe phases for 0.7 $\lesssim$ $\Delta$ $\lesssim$ 1~\cite{zhu2018topography}.

Such Kitaev-like BD terms arise naturally in magnets with strong spin-orbit coupling~\cite{li2017kitaev,takagi2019concept} when the total multiplet of a magnetic ion is projected onto the low-energy pseudo-$S$ = 1/2 model. However, the experimental realizations of such terms remain very rare~\cite{avers2021finger,kim2023bond}. Rare-earth delafossites represent a promising playground for the studies of TLAF because of their distortion-free crystal structure with well-separated triangular layers~\cite{liu2018rare,Baenitz2018}. Yb-based materials are known to potentially possess QSL~\cite{ding2019gapless,bordelon2019field,Xing2019,dai2020spinon} or the 120$^{\circ}$ ordered states~\cite{xie2023complete,Scheie2021}, while Ce~\cite{Bastie2020, Kulbakov2021, avdoshenko2022spin} and Er~\cite{Xing2021KErSe2,ding2023stripe}-based materials order into the stripe phases [Fig.~\ref{Fig1}(b)] because of the BD terms. However, there has been no report on the anomalous spin dynamics caused by the BD terms in these materials until now.

Here, we focus on \ccs\ that orders magnetically into the stripe-$yz$ phase below $T_{\rm N}$ = 0.35~K~\cite{Xie2023stripe}. A combination of crystalline electric field and spin-orbit coupling splits the $J = 5/2$ multiplet of Ce$^{3+}$ into three well-separated Kramers doublets, with the ground state doublet corresponding to an effective $S = 1/2$~\cite{li2016hidden,Xie2023stripe}. Application of the magnetic field, $\mathbf{B}~||~\mathbf{a}$, suppresses the stripe order at a potential quantum critical point and stabilizes the field-polarized (FP) state [Fig.~\ref{Fig1}(e)].
We collect inelastic neutron scattering (INS) spectra at all relevant fields and our results reveal the anomalous spin excitations in the ordered stripe and FP phases.
Using these data we refine the spin Hamiltonian and establish that the BD terms are responsible for the stabilization of the stripe-$yz$ order. Our INS spectra and exact diagonalization (ED) calculations demonstrate a large broadening of the spectra at intermediate fields and complete breakdown of magnons at the $\Gamma$ point of the Brillouin zone (BZ) just below the critical field ($B_{\rm c}$ = 3.86~T).
We also find strong repulsion between the single magnon branch and the two-magnon continuum (TMC) that is best visible at the Y point of the BZ.
Our results unravel the crucial role of the magnon-magnon interaction induced by the BD terms in modifying the excitation spectrum of the TLAF with a magnetically ordered ground state.

\begin{figure*}[tb]
\center{\includegraphics[width=.95\linewidth]{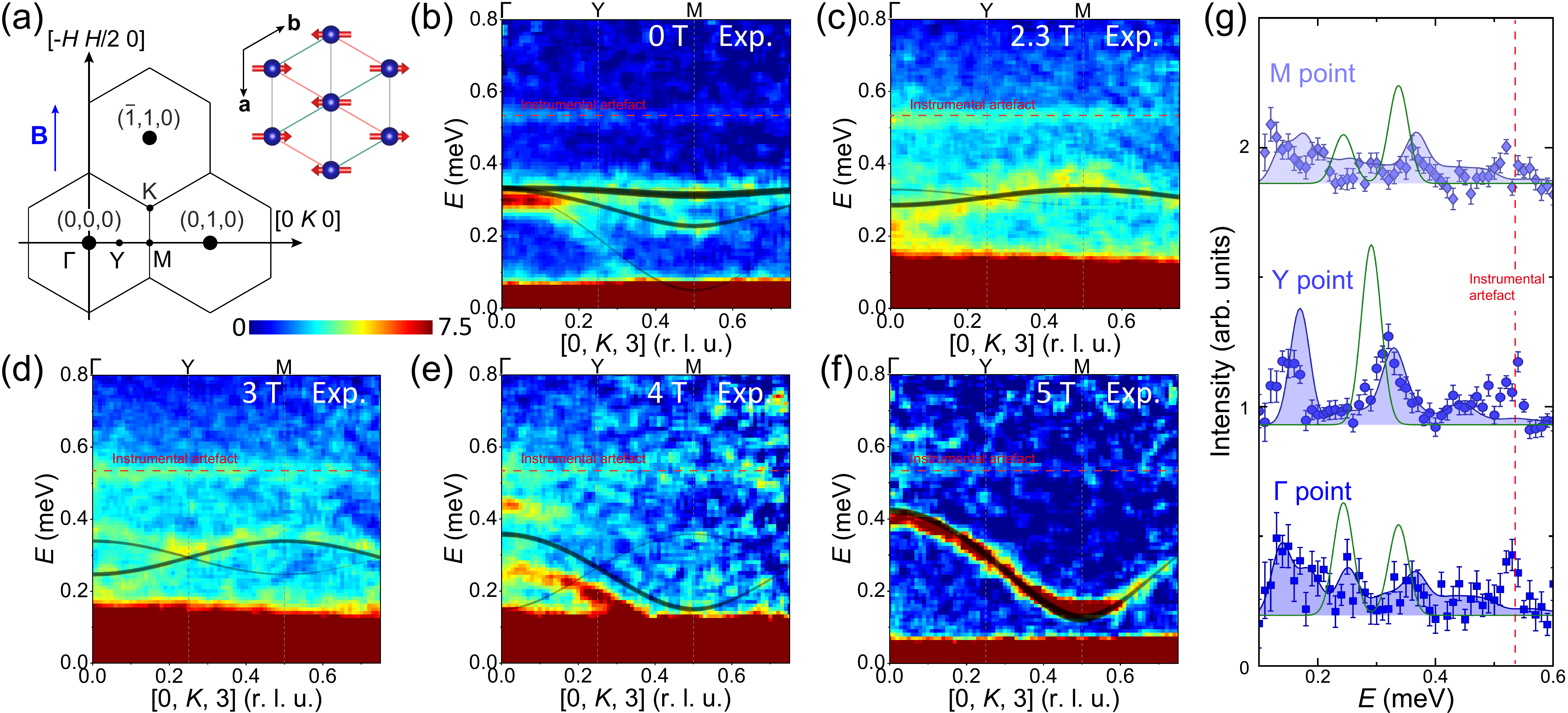}}
  \caption{
  (a)~Left: representation of the BZ in the ($H$$K$0) plane. The blue arrow indicates the magnetic field direction. Right: projection of the Ce$^{3+}$ moments on the $ab$ plane at zero field.
  (b)--(f)~Comparisons of the experimental spectra with results of the LSWT calculations at different magnetic fields. All INS data were taken  at $T$ = 70~mK and integrated by $\pm2.5$~r.l.u. along [0~0~$L$] and $\pm{}0.1$~r.l.u. along [$H$~$-H$/2~0] directions. The black lines on the top of color maps are the mode positions and intensities given by LSWT.
  (g)~Comparison between the INS data and calculations at different $\mathbf{Q}$ points and $B$ = 3~T~\cite{Background}. The shadow areas are ED and the green curves are the LSWT results, respectively. The red dashed lines in panels (b)-(g) indicate the instrumental artefact $\approx$0.54~meV.
  }
  \label{Fig2}
  \vspace{-12pt}
\end{figure*}

\textit{Spin dynamics in \ccs.}--- \ccs\ orders into the two-sublattice stripe-$yz$ phase [Fig.~\ref{Fig2}(a)] which breaks the $C_3$ lattice symmetry~\cite{Xie2023stripe}. The corresponding magnetic BZ is nested in the crystallographic one [Fig.~\ref{Fig1}(c)]. As a consequence, the otherwise-equivalent high-symmetry points of the BZ become nonequivalent. Below we label the two M points associated with the ordering wave vector as M and the four others as M$'$ [Fig.~\ref{Fig1}(c)].

Figure~\ref{Fig2}(b) shows the INS spectrum at zero field. Its primary feature is a sharp intense excitation at the $\Gamma$ point. Away from the zone center, \ccs\ exhibits three spin-wave branches: one nearly gapless acoustic and two weakly dispersive optical modes. The canted stripe-$yz$ order has two magnetic sublattices and, in the spin-wave approximation, \ccs\ should feature two magnon modes at $B$ $<$ $B_{\rm c}$.
However, in zero field \ccs\ is in a multidomain state, and, therefore, the INS spectrum consists of the superposition of three domains, $I(\mathbf{Q},\hbar\omega)$=$I(\mathbf{Q}_{\Gamma\xrightarrow{}\rm{}M},\hbar\omega)$+$2I(\mathbf{Q}_{\Gamma\xrightarrow{}\rm{}M'},\hbar\omega)$.
This increases the number of modes up to six, but the two pairs of these modes, originating from the domains ordered at the M$'$ points, are exactly degenerate along the $\Gamma$$\xrightarrow{}\rm$M path.

Application of the magnetic field selects one domain with the ordering wavevector $\mathbf{q}$$\perp$$\mathbf{B}$, and the spectra obtained at 1.4~(see Supplemental Material (SM)~\cite{SI}) and  2.3~T [Fig.~\ref{Fig2}(c)] yield a weakly dispersive mode around 0.35~meV~\footnote{The second mode is primarily polarized along the $\mathbf{c}$ axis, and its intensity is strongly suppressed due to the anisotropy of the $g$ factor.}.
At 3~T [Fig.~\ref{Fig2}(d)] the spectrum exhibits qualitative modifications: the mode becomes more dispersive and the excitation at the $\Gamma$ point strongly broadens in energy [Fig.~\ref{Fig2}(g)]. At 4~T the spectrum demonstrates a low-energy dispersive mode and an arch-shaped resonancelike feature at $E$$\approx$0.45~meV, which is localized at a close proximity of the $\Gamma$ point [Fig.~\ref{Fig2}(e)]. The intensity distribution over the low-energy mode is not homogeneous and it weakens close to the $\Gamma$ point, just below the high-energy arch. Further increase of the field stabilizes the FP state and the spectrum exhibits a sharp mode with a cosinelike dispersion~[Fig.~\ref{Fig2}(f)]. We note that in the high-field regime, $B$ $\geq$ 5~T, the observed excitations are resolution limited, but broaden considerably in the lower fields~\cite{SI}.

Now we turn to the determination of the Hamiltonian parameters. The spin Hamiltonian for the NN triangular lattice reads as~\cite{li2015rare,Liyaodong2016,Liyaodong2018,zhu2018topography,maksimov2019anisotropic}:
\begin{align}  \label{Hamiltonian}
\mathcal{H} = \sum_{\langle ij \rangle} \mathbf{S}_i^{\mathrm{T}} \hat{\bf J}_{ij} \mathbf{S}_j - \mu_B g_{ab} B \sum_i{S}^x_i
\end{align}
where $\mu_{\rm B}$ is the Bohr magneton, $g_{ab}$ is the in-plane $g$ factor, $B$ is the magnetic field, and $\hat{\bf J}_{ij}$ is the exchange matrix, defined as
\begin{align}  \label{Exchange_matrix}
\hat{\bf J}_{ij} =	\left(
			\begin{array}{ccc}
				J+2J_{\pm\pm}    &0			    &0			\\	
				0			    &J-2J_{\pm\pm}   &J_{z\pm}	\\
				0			    &J_{z\pm}	    &\Delta{}J	\\
    \end{array}
    \right),
\end{align}
for the bond along the $\mathbf{a}$ axis, which transforms according to the lattice symmetry for the other bonds~\cite{maksimov2019anisotropic}. The Hamiltonian~\eqref{Hamiltonian} has five independent variables, which we determine by fitting the dispersion in the high-field regime and using calculations of the low-field spectral response using ED as is detailed in the SM~\cite{SI}.  Our results safely exclude considerable next-NN interactions, and the best fit yields $J$ = 72.5~$\mu$eV, $\Delta$ = 0.25, $J_{\pm \pm}/J$ = 0.52, $J_{z \pm}/J$ = 0.41, and $g_{ab}$ = 1.77~\footnote{Equivalently, these parameters can be written using $J$-$K$-$\Gamma$-$\Gamma'$ notations~\cite{maksimov2019anisotropic, maksimov2022Erratum, maksimov2022easy}:  $J = -82.5~\mu$eV, $K = -33.3~\mu$eV, $\Gamma = -65.5~\mu$eV and $\Gamma' = 14~\mu$eV. This indicates that $J$ and $\Gamma$ are the leading terms and the Kitaev interaction $K$ is of the same order as the leading terms.}\footnote{As detailed in the SM, these values of the couplings should be understood as a representative set of couplings rather than an ``exact'' estimate.}.
These parameters put \ccs\ deep into the stripe-$yz$ phase of the general phase diagram of the TLAF [Fig.~\ref{Fig1}(a)] and demonstrate that the stripe order is stabilized by a combination of the strong BD terms.

We note that the linear spin-wave theory (LSWT) and symmetry consideration~\cite{karaki2023efficient} suggest the Dirac crossing in the magnon spectrum at the Y points [Fig.~\ref{Fig1}(d)]. While the presence of the magnon-magnon interaction may suggest some exotic interplay of it with the Dirac topology~\cite{mcclarty2019non, mook2021interaction}, our results presented below are unable to support such a scenario.

The LSWT calculations of the spin wave for each field are summarized in Figs.~\ref{Fig2}(b)-\ref{Fig2}(f). Clearly, LSWT captures perfectly the high-field data and reproduces some features of the \textit{dispersion} in the low-field regime, $B$$\leq$ 2.3 ~T, with semiquantitative accuracy. However, accurate comparison of the intensities clearly demonstrates that at intermediate field regime 0$<$ $B$ $<$ $B_{\rm c}$, the LSWT simulations do not provide an adequate description of the observed data despite the presence of robust magnetic order~\cite{Xie2023stripe}. To show that, in Fig.~\ref{Fig2}(g) we plot the INS data taken at 3~T at the $\Gamma$, Y, and M points together with LSWT and ED results. According to the LSWT, the spectra at the $\Gamma$ and M points contain two sharp modes in a clear contrast with the INS data, which exhibit a very broad response with only a weak hint of the LSWT-predicted modes, indicating magnon breakdown due to strong magnon-magnon interactions. Moreover, the LSWT spectrum at the Y point features a single mode due to topological protection, while our data clearly show the two-peak structure, providing strong evidence that a more sophisticated treatment is required.

To provide an accurate description of the observed INS data, we utilize unbiased numerical ED calculations on a finite system size up to $N$ = 32 spins. We note that the spectral function computed with ED and LSWT are converted to the INS cross section taking into account all relevant factors~\cite{SI}. This allows us to make quantitative comparisons between INS and ED at three high-symmetry points ($\Gamma$, Y and M) of the BZ.

\begin{figure}[tb]
\center{\includegraphics[width=.95\linewidth]{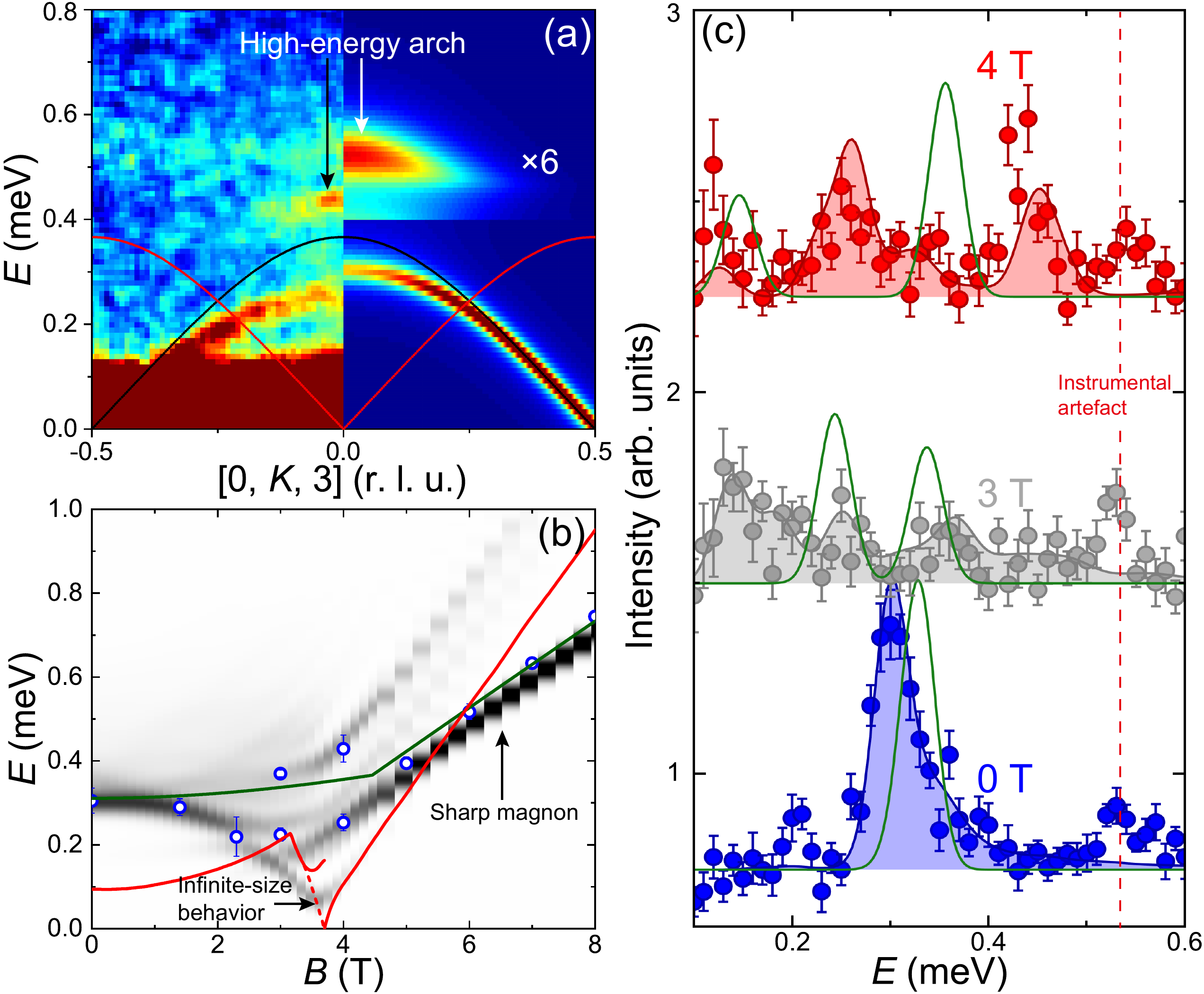}}
  \caption{
    (a)~The INS spectrum of \ccs\ collected at 4~T (left) compared with nonlinear SWT calculations (right). The black line is the LSWT dispersion and the red line is the bottom of the TMC at 4~T. The arrows point to the high-energy ``arch'' within the TMC. The intensity of the upper part of the calculated spectrum ($E$ $>$ 0.4~meV) is multiplied by $\times$6 for better visual clarity.
    (b)~Calculated dynamical structure factor at the $\Gamma$ point and different fields obtained using ED. The blue open circles are experimental results. The dark green line is the LSWT results. The red line is the lower boundary of the TMC.
    (c)~Comparisons of the experimental and the calculated spin excitation spectra at the $\Gamma$ point~\cite{Background}.
    The shadow areas are ED calculations, and the green curves are LSWT results.
  }
  \label{Fig3}
  \vspace{-12pt}
\end{figure}

First, we consider the spectrum measured at 4~T that contains the high-energy arch near the $\Gamma$ point [Fig.~\ref{Fig3}(a)]. This field is just above $B_{\rm c}$~\cite{Xie2023stripe}, and, based on the LSWT, one would expect a resolution-limited magnon branch. Indeed, our INS data show the presence of a sharp low-energy magnon branch which, however, strongly broadens close to the $\Gamma$ point. The INS spectrum can be reproduced by the nonlinear SWT calculation with a semiquantitative accuracy. It reproduces the high-energy arch around the $\Gamma$ point that can be observed clearly, despite its weaker intensity [Fig.~\ref{Fig3}(a)].
Our Fig.~\ref{Fig3}(b) shows the single-magnon dispersion and the lower boundary of the TMC versus field obtained by ED~\cite{SI}. One can see that the broadening of the single-magnon branch takes place when the magnon branch enters the TMC. Such an overlap makes the decay of the single magnon kinematically allowed below $\approx$5.5~T and explains the momentum-dependent broadening of the magnon mode. Our ED calculations clearly support this scenario and also reproduce the formation of the high-energy resonance feature, which we interpret as a large density of states within the TMC.

\begin{figure}[tb]
\center{\includegraphics[width=1\linewidth]{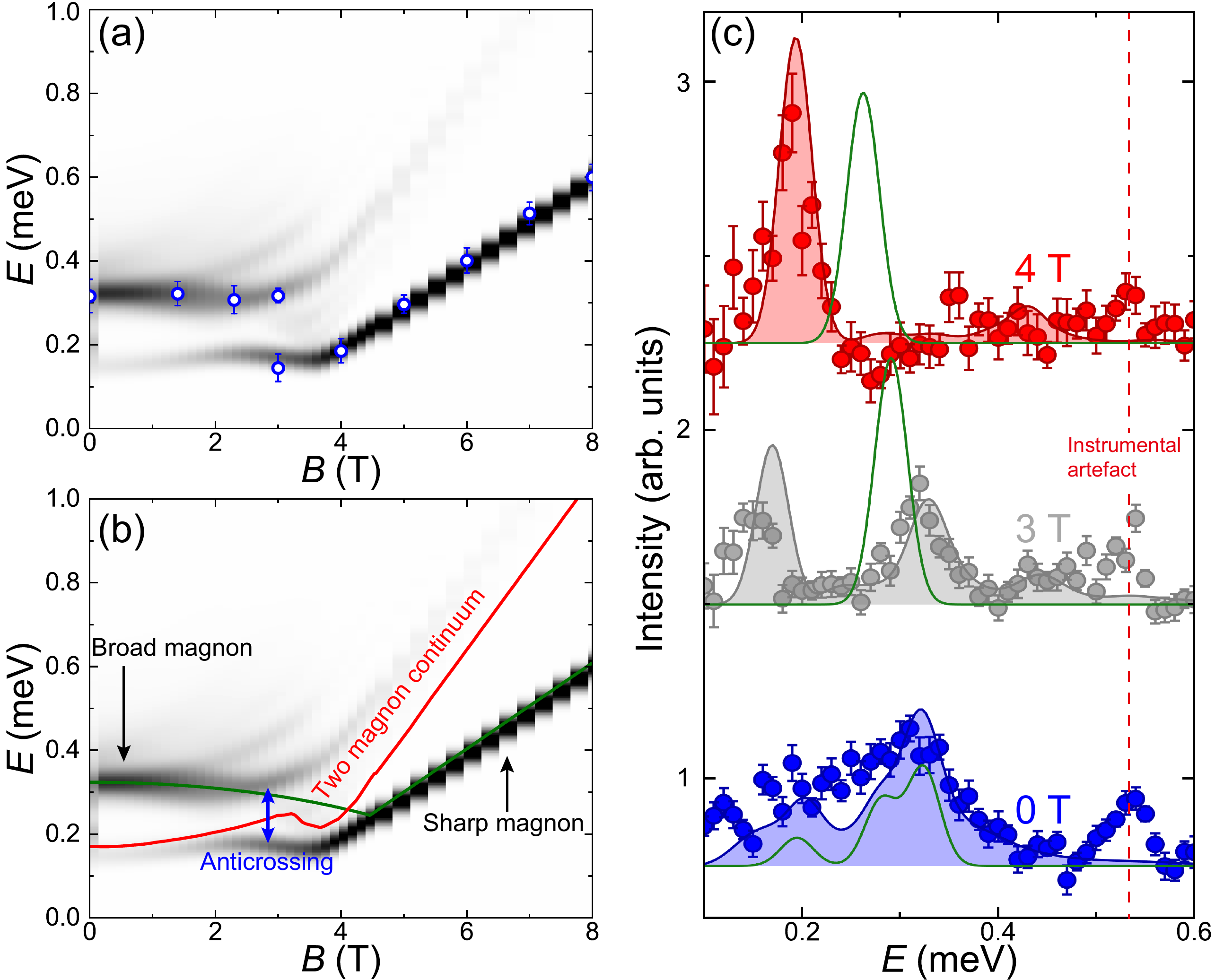}}
  \caption{
  (a)~Dynamical structure factor of \ccs\ calculated by ED. The blue open circles are experimental results.
  (b)~A replica of the ED calculations in panel (a) with different regions indicated with single magnon and TMC. The dark green line is LSWT calculations. The red line is the lower boundary of the TMC. The blue arrow emphasizes the anticrossing behavior between the single-magnon branch and the TMC.
  (c)~Comparisons of the experimental and the calculated spectra at Y point~\cite{Background}.
  The shadow areas are ED calculations, and the green curves are LSWT results.
  }
  \label{Fig4}
  \vspace{-12pt}
\end{figure}

At $B_{\rm c}$, the TMC extends to vanishing energy at the $\Gamma$ point~\footnote{In ED, this is only true in the thermodynamic limit.}. This causes a strong broadening of the magnetic signal at the spectra collected at 2.3 and 3~T where the experimental spectral weight is smoothly distributed between 0.1 and 0.4~meV indicating a complete breakdown of the single-magnon excitation.
In contrast, a relatively sharp Lorentzian-shaped mode is formed at lower fields, see Fig.~\ref{Fig3}(b). While the LSWT captures some parts of the spectral response at 0~T, it fails in the intermediate field regime, while ED reproduces our INS data with a high accuracy at all relevant fields, see Fig.~\ref{Fig3}(c). We speculate that the failure of the LSWT at intermediate fields is expected because the spectra are subject to stronger quantum effects for the canted, noncollinear spin configuration.
That allows for a stronger coupling between the longitudinal and transverse spin excitations via the three-magnon processes, opening up additional channels for magnon decays~\cite{zhitomirsky2013colloquium}.

Now we turn to the description of the data collected at the Y point.
As we discussed above, the LSWT predicts a topologically protected Dirac point in the stripe-$yz$ phase.
Thus, the spectrum at the Y point should feature one mode at every field above zero. However, the magnon-magnon interaction induces strong modifications of the spectrum. Figures~\ref{Fig4}(a) and \ref{Fig4}(b) show the ED-calculated INS spectra at different fields. In the FP regime, $B$ $>$ $B_{\rm c}$, it shows a sharp mode that exhibits a nonmonotonic behavior in the ordered phase $B$ $<$ $B_{\rm c}$: first it bends up just below $B_{\rm c}$, qualitatively similar to the predictions of LSWT. However, at $B$ $\approx$ 2.5~T it twists in the opposite direction. Moreover, most of the spectral weight at low fields is transferred from this mode to the high-energy continuum at 0.3--0.4~meV. Notably, the position of the intense excitation at low fields agrees quantitatively with the LSWT predictions.

To rationalize the origin of such a behavior we calculate the lower boundary of the TMC at the Y point using ED, shown by the red line in Fig.~\ref{Fig4}(b). The low-energy mode is located just below the lower edge of the TMC. Such a behavior is caused by a repulsion between the single-magnon branch and the continuum in the avoided quasiparticle decay scenario~\cite{Plumb2016Quasiparticle,Verresen2019Avoided}. Our INS data support this scenario and show two well-separated INS peaks at 3~T, see~Fig.~\ref{Fig4}(c). In our setup, we can control and track the strength of the repulsion by applying magnetic field, in contrast to that in~Ref.~\cite{Verresen2019Avoided}, where the repulsion is manifested as a function of the momentum transfer.

\textit{Discussion and Conclusion.}--- The interaction between a quasiparticle branch and a continuum has two opposite regimes: when the interaction is weak, the quasiparticle acquires a finite lifetime where the mode enters the continuum. In the case of strong interactions, a level repulsion scenario can take place, meaning that the quasiparticle dispersion will repel from the continuum remaining sharp in energy.
Our spectroscopic study of \ccs\ demonstrates the presence of both effects which arise as a consequence of interactions between the single-magnon branches and the multimagnon continuum. This interaction has substantial matrix elements because of the sizable bond-dependent terms, $J_{\pm\pm}$ and $J_{z\pm}$, in the spin Hamiltonian and, furthermore, its effect can be controlled by a magnetic field. Specifically, using the 4~T dataset we demonstrate that once the single-magnon branch enters the TMC, it exhibits downward renormalization and acquires a finite lifetime, in agreement with the weak-interacting scenario, while part of the spectral weight is transferred to the TMC. The spectra at the $\Gamma$ point at 2.3~T and 3~T show a very broad response indicating a complete breakdown of the single-magnon excitation which, however, is almost reestablished in the collinear zero-field phase.

The data collected at the Y point demonstrate anticrossinglike behavior, where the single-magnon branch is repelled by the lower boundary of the TMC, remaining relatively sharp in energy. This behavior is induced by a strong interaction between the magnon branch and the continuum, which was observed previously in the excitation spectrum of liquid $^4$He and in the TLAF Heisenberg magnet~\cite{Verresen2019Avoided,donnelly1981specific, glyde1998excitations,Godfrin2021,Macdougal2021}. Our results are essential for the understanding of the many-body spin dynamics in materials that combine geometrical and bond frustration.

\begin{acknowledgments}
\textit{Acknowledgements.}---
We thank Dr. Hongtao Liu (Instrument Analysis \&\ Research Center, Sun Yat-sen University) for the assistance with LA-ICP-TOF measurements and data evaluations.
We thank Dr.~Jong Keum for the help with the x-ray Laue and x-ray diffraction measurements, Dr.~Feng Ye for assistance with single-crystal x-ray diffraction measurements, N.~D.~Andriushin for the help with figure preparations, Dr.~J\"org Sichelschmidt for the attempts of the ESR measurements and the helpful discussion.
Work at Sun Yat-sen University was supported by the National Natural Science Foundation of China (Grant No. 12304187), the open research fund of Songshan Lake Materials Laboratory (Grant No. 2023SLABFN30), the Guangzhou Basic and Applied Basic Research Funds (Grant No. 2024A04J4024), and the Fundamental Research Funds for the Central Universities, Sun Yat-sen University (Grant No. 23qnpy57).
Work at Oak Ridge National Laboratory (ORNL) was supported by the U.S. Department of Energy (DOE), Office of Science, Basic Energy Sciences, Materials Science and Engineering Division. This research used resources at the Spallation Neutron Source, a DOE Office of Science User Facility operated by the Oak Ridge National Laboratory.
X-ray Laue and XRD measurements were conducted at the Center for Nanophase Materials Sciences (CNMS) (CNMS2019-R18) at ORNL, which is a DOE Office of Science User Facility. The work of A. L. C. on the analytical SWT was supported by the U.S. Department of Energy, Office of Science, Basic Energy Sciences under Award No. DE-SC0021221.
\end{acknowledgments}


\begin{thebibliography}{70}
\bibitem [{Bac()}]{Background}%
  \BibitemOpen
  \href@noop {} {}\bibinfo {howpublished} {The background was subtracted from
  the INS data as detailed in Sec.~S1B of SM~\cite{SI}.}\BibitemShut {Stop}%
\end{thebibliography}
\end{document}


\title{Supplemental Material\\ Quantum Spin Dynamics Due to Strong Kitaev Interactions in the Triangular-Lattice Antiferromagnet~\ccs}

\author{Tao Xie}
\thanks{Corresponding author: xiet69@mail.sysu.edu.cn}
\affiliation{Center for Neutron Science and Technology, Guangdong Provincial Key Laboratory of Magnetoelectric Physics and Devices, School of Physics, Sun Yat-sen University, Guangzhou, Guangdong 510275, China}
\affiliation{Neutron Scattering Division, Oak Ridge National Laboratory, Oak Ridge, Tennessee 37831, USA}
\author{S. Gozel}
\affiliation{Laboratory for Theoretical and Computational Physics, Paul Scherrer Institute, CH-5232 Villigen-PSI, Switzerland}
\author{Jie Xing}
\affiliation{Materials Science and Technology Division, Oak Ridge National Laboratory, Oak Ridge, Tennessee 37831, USA}
\author{N. Zhao}
\affiliation{Department of Physics, Southern University of Science and Technology, Shenzhen, Guangdong, 518055, China.}
\author{S. M. Avdoshenko}
\affiliation{Leibniz-Institut f\"ur Festk\"orper- und Werkstoffforschung (IFW Dresden), Helmholtzstra{\ss}e 20, 01069
Dresden, Germany}
\author{L. Wu}
\affiliation{Department of Physics \&\ Academy for Advanced Interdisciplinary Studies, Southern University of Science \&\ Technology, Shenzhen, Guangdong, 518055, China}
\author{Athena~S.~Sefat}
\affiliation{Materials Science and Technology Division, Oak Ridge National Laboratory, Oak Ridge, Tennessee 37831, USA}
\author{A. L. Chernyshev}
\affiliation{Department of Physics and Astronomy, University of California, Irvine, California 92697, USA}
\author{A. M. L\"auchli}
\affiliation{Laboratory for Theoretical and Computational Physics, Paul Scherrer Institute, CH-5232 Villigen-PSI, Switzerland}
\affiliation{Institute of Physics, Ecole Polytechnique F\'ed\'erale de Lausanne (EPFL), CH-1015 Lausanne, Switzerland}
\author{A.~Podlesnyak}
\affiliation{Neutron Scattering Division, Oak Ridge National Laboratory, Oak Ridge, Tennessee 37831, USA}
\author{S.~E.~Nikitin}
\thanks{Corresponding author: stanislav.nikitin@psi.ch}
\affiliation{Laboratory for Neutron Scattering and Imaging, Paul Scherrer Institut, CH-5232 Villigen-PSI, Switzerland}

\maketitle

Section~\ref{smsec::ins_experiment} presents details on the sample preparation and inelastic neutron scattering (INS) measurements. Section~\ref{smsec::NPD} details determination of the magnetic structure from the neutron powder diffraction data. Section~\ref{smsec::ed} shortly discusses the exact diagonalization (ED) technique. Details on the procedure used to fit the coupling constants of the model are presented in Sec.~\ref{smsec::fit_couplings}. Section~\ref{smsec::lswt} presents the linear spin wave theory (LSWT) magnon dispersion and two-magnon threshold across the Brillouin zone (BZ) at intermediate fields. Section~\ref{smsec::stripeyz} briefly discusses magnetic structure and the influence of the out-of-plane tilt on the magnon topology. Section~\ref{Sec_NLSWT} discusses a non-linear spin wave theory analysis.

\section{Inelastic neutron scattering experiment}\label{smsec::ins_experiment}
\subsection{Details of experiments}
The high-quality single crystals of \ccs\ were grown using a flux method~\cite{Xing20191}. We co-aligned about 90 single crystals in the $(0~KL)$ scattering plane on thin copper plates to get a mosaic sample with mass $\sim$0.4~g and mosaicity $\sim$2\degree. The neutron scattering data were collected at the time-of-flight Cold Neutron Chopper Spectrometer (CNCS)~\cite{CNCS1,CNCS2} at the Spallation Neutron Source, Oak Ridge National Laboratory. We performed two experiments, named Exp.1 and Exp.2, using 5~T and 8~T cryomagnets equipped with a dilution refrigerator insuring sample temperature below 0.1~K. In the first experiment we collected two datasets at 0 and 5~T. In the second experiment we have measured the data at 1.4, 2.3, 3, 4, 6, 7 and 8~T. All data used to analyse details of the low-energy spin dynamics were collected with $E_{\rm i} = 1.55$~meV (energy resolution is 0.035~meV at the elastic line). Several additional dataset were collected in the paramagnetic phase at $T = 12$~K and at $T = 0.07$~K with $E_{\rm i} = 3.32$~meV.

\begin{figure}[tb]
    \includegraphics[width=.45\textwidth]{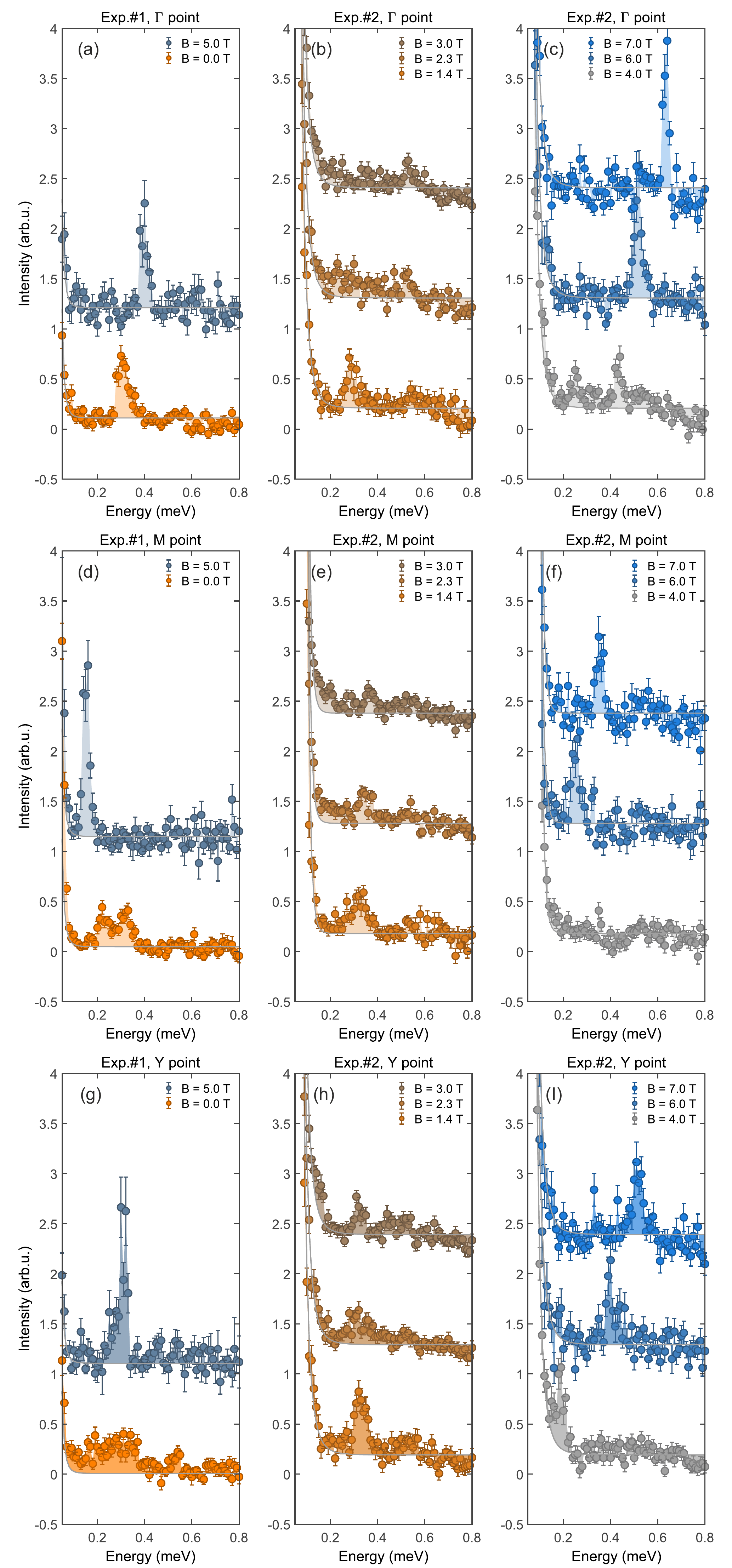}
    \caption{~Raw INS spectra at $\Gamma$, Y and M points taken at $T = 70$~mK at different fields as indicated in the legends. Data in panels (a,d,g) were collected in the first experiment, while the data in panels (b,c,e,f,h,i) -- during the second one. Grey lines in every panel represent a background signal fitted globally for each experiment and $\mathbf{Q}$ point. The curves are shifted vertically for clarity.}
    \label{Fig_SI_all_INS_cuts}
\end{figure}

During the measurements, the vertical magnetic field was applied along the $\left[ 1 \, -0.5 \, 0 \right]$ direction in $ab$ plane.
We employed the software packages~\textsc{MantidPlot}~\cite{Mantid} and~\textsc{Horace}~\cite{Horace} for data reduction and analysis. The linear spin wave theory simulations were performed with~\textsc{SpinW}~\cite{spinw}.

\subsection{Background model}\label{sec::bg_INS}
In Figs.~2(g), 3(c) and 4(c) of the main text we present the representative background-subtracted INS spectra at different high-symmetry points.
The background was modeled using a power law function:
\begin{align}
    \mathrm{BG}(E) = a + b E^{-c}
    \label{eq::bg}
\end{align}
where $a, b$ and $c$ are empirical coefficients. These coefficients were fitted individually for different $\mathbf{Q}$ points and for different experiments due to the background difference from the different sample environments in the two experiments. But for a given experiment and $\mathbf{Q}$ point the coefficients were fitted globally taking into account all fields measured at this setup (for instance, $a,b$ and $c$ are the same for 0 and 5~T curves at M point).
Figure~\ref{Fig_SI_all_INS_cuts} shows the constant-$\mathbf{Q}$ cuts taken at every field at $\Gamma$, M and Y points of BZ. Grey lines below each curve show an estimated background contribution and the shaded areas represent the signal. The INS signal shown in Figs.~2(g), 3(c) and 4(c) of the main text is the difference between the raw INS data and the fitted BG contribution.

\subsection{Summary of the INS data}\label{sec::summary_INS}
\begin{figure*}[tb]
    \center{\includegraphics[width=.99\textwidth]{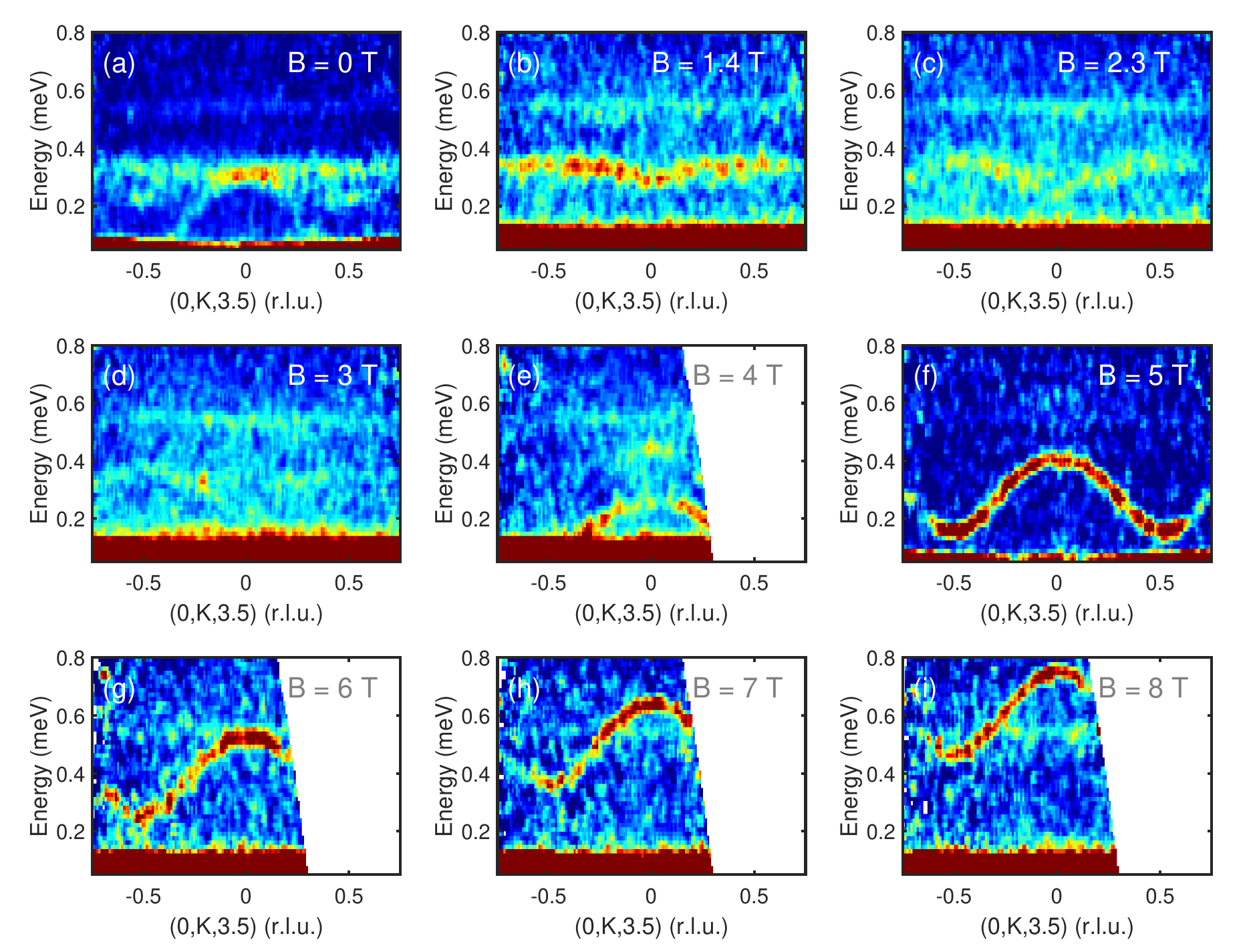}}
    \caption{INS data taken at $T = 70$~mK and different magnetic fields as indicated in each panel. The data are integrated by $\pm1.5$~r.l.u. along [0~0~$L$] and $\pm{}0.05$~r.l.u. along [$H$~$-H$/2~0].
    }
    \label{Fig_SM_all_INS}
\end{figure*}

In the main text we show only the spectra collected at 0, 2.3, 3, 4 and 5~T. In this section we provide the full overview of our low-temperature INS data and all spectra are summarized in Fig.~\ref{Fig_SM_all_INS}. We note that the spectra have different statistics. The high-field data at $B \geq\ 5$~T [Fig.~\ref{Fig_SM_all_INS}(f-i)] exhibit strong sharp magnon mode and thus each dataset was counted for only a short time, $\approx 4$~h. In contrast, the low- and especially intermediate-field datasets exhibit weak, broad magnetic response and therefore have been measured for much longer time, up to 24~h for the 2.3 and 3~T datasets.

The spectrum collected at $B = 3$~T demonstrates almost complete breakdown of the magnon quasiparticles and even distribution of the spectral weight almost everywhere in $\mathbf{Q}$ implying a possible presence of quasi-elastic scattering as a manifestation of critical fluctuations of the order parameter that can in principle be observed in INS. Indeed, constant-$\mathbf{Q}$ cuts taken at 3~T [Figs.~\ref{Fig_SI_all_INS_cuts}(b,e,h)] show a small spectral weight close to the elastic line. But unfortunately, due to high background contribution we cannot resolve a clear quasi-elastic signal nor quantify the lifetime of such fluctuation and additional measurements with a larger sample are required to address this question.

\subsection{Instrumental artefact at 0.54~meV}\label{sec::artefact_INS}
\begin{figure}[tb]
    \center{\includegraphics[width=.99\linewidth]{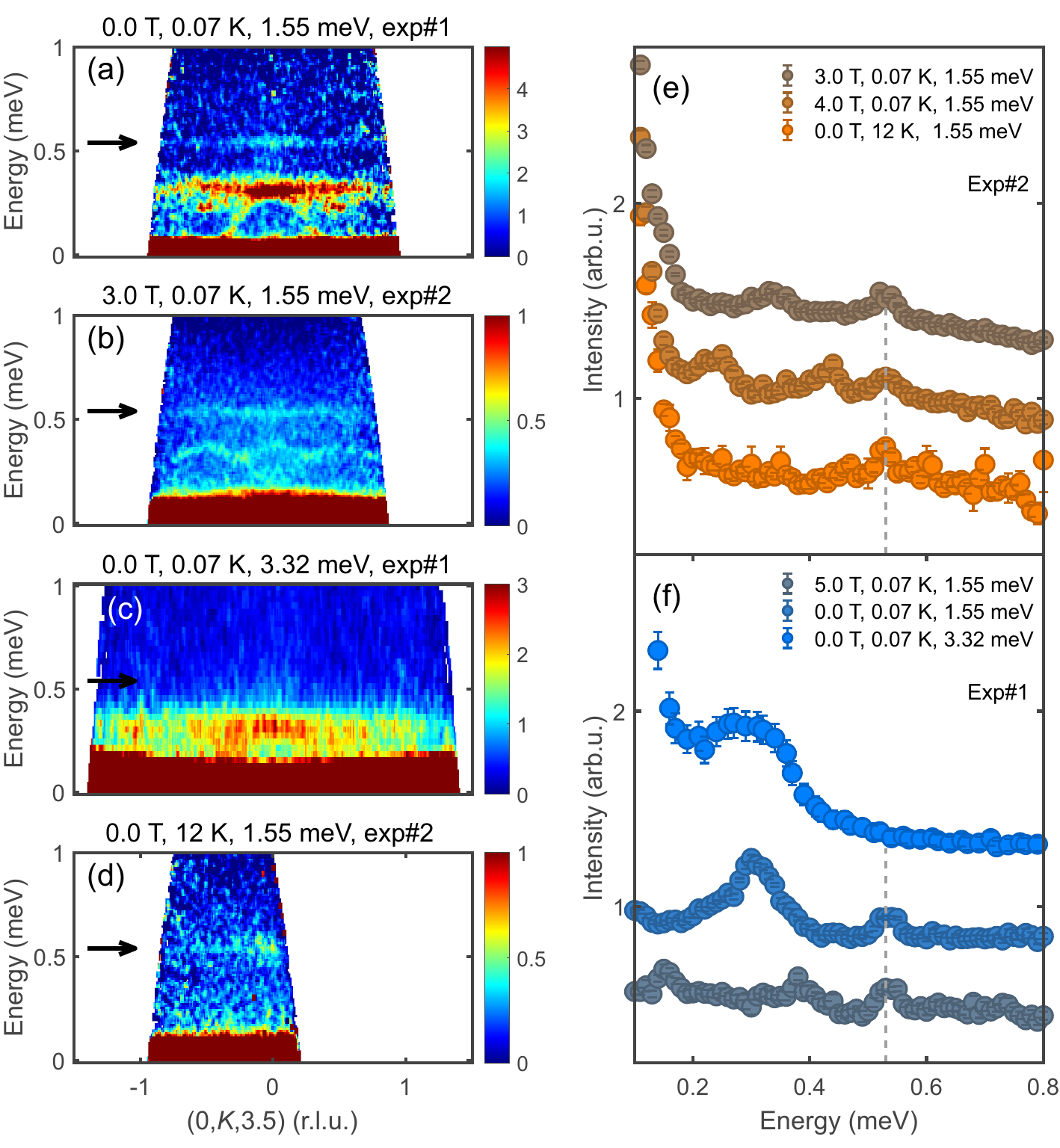}}
    \caption{~Instrumental artefact at 0.54~meV.
    (a)~Spectrum collected at $T = 0.07$~K, $B = 0$~T and $E_{\rm i} = 1.55$~meV using 5~T cryomagnet.
    (b)~Spectrum collected at $T = 0.07$~K, $B = 3$~T and $E_{\rm i} = 1.55$~meV using 8~T cryomagnet.
    (c)~Spectrum collected at $T = 0.07$~K, $B = 0$~T and $E_{\rm i} = 3.32$~meV  using 5~T cryomagnet. Note that the horizontal stripe is absent.
    (d)~Spectrum collected at $T = 12$~K,  $B = 0$~T and $E_{\rm i} = 1.55$~meV  using 8~T cryomagnet.
    (e,f)~One-dimensional INS spectra integrated over $K = [-1,1]$ and $L = [1,6]$ r.l.u. Data in panels (e) and (f) were taken in 8~T and 5~T cryomagnets respectively (Exp$\#$2 and Exp$\#$1).
    Blue curve collected with $E_{\rm i} = 3.32$~meV was scaled by $\times 0.25$ for visual comparison. The curves are shifted vertically for clarity.
    Dotted grey lines in panels (e,f) and the black arrows panels (a-d) indicate the position of the 0.54~meV artefact.
    }
    \label{Fig_SM_artefact_INS}
\end{figure}

As can be seen in Fig.~\ref{Fig_SM_all_INS}, all spectra collected with $E_{\rm i} = 1.55$~meV show a weak dispersionless line at 0.54~meV. If this feature is a part of magnetic (intrinsic) signal, it should demonstrate some dependence on external parameters such as temperature or magnetic field and, more importantly, be present in the datasets collected under identical conditions but with different incident neutron energies.
In Fig.~\ref{Fig_SM_artefact_INS} we summarize the two evidences of the opposite.
First, in Figs.~\ref{Fig_SM_artefact_INS}(a) and (b) we show the spectra taken at 0 and 3~T at 0.1~K (using 5 and 8~T cryomagnets, respectively). Clearly, the line is present at both datasets and does not change with magnetic field. Moreover, the same feature can also be seen clearly in the dataset measured at 12~K [Fig.~\ref{Fig_SM_artefact_INS}(d), 8~T cryomagnet], well above the magnetic transition temperature. Note that this dataset was collected with considerably poorer statistics compared to the low-temperature spectra and thus, the artifact has worse visual contrast.

As the next step, we made several one-dimensional energy cuts on the representative datasets, the spectra taken using 8~T and 5~T cryomagnets are summarized in Figs.~\ref{Fig_SM_artefact_INS}(e) and \ref{Fig_SM_artefact_INS}(f) respectively. The feature at 0.54~meV appears to be field- and temperature-independent for all  datasets collected with $E_{\rm i} = 1.55$~meV. As we commented above, a relatively poor contrast in the 12~K spectrum is because of the short counting time, yet the feature remains resolvable.

Second, the dataset collected with $E_{\rm i} = 3.32$~meV at zero field at 0.1~K does not show any inelastic signal at 0.54~meV as can be seen clearly in Figs.~\ref{Fig_SM_artefact_INS}(c,f). We note that neutron flux of CNCS is much higher for $E_{\rm i} = 3.32$~meV as compared to $E_{\rm i} = 1.55$~meV setting (approximately by a factor of 4). It means that 3.32~meV should demonstrate a resolvable signal at 0.54~meV, if it would have magnetic (intrinsic) origin, which contradict our results. The absence of the 0.54~meV feature in 3.32~meV dataset can be seen most clearly in the one-dimensional cut, Fig.~\ref{Fig_SM_PND}(f). The spectrum demonstrates strong magnetic scattering at $E < 0.4$~meV and very monotonic background signal at $E > 0.45$~meV without any hint of an excitation at 0.54~meV. Thus, we show that the signal at 0.54~meV is field- and temperature independent and disappears when we change the incident neutron energy. Thus, we conclude that the dispersionless stripe observed at 0.54~meV has extrinsic origin. Most likely, this signal appears due to multiple scattering and is specific for $E_{\rm i} = 1.55$~meV setting of the CNCS instrument.

\section{Magnetic structure determined from the powder neutron diffraction}\label{smsec::NPD}
To determine the magnetic structure of \ccs\ we performed powder neutron diffraction measurements using HB-2A (POWDER) diffractometer at High Flux Isotope Reactor, Oak Ridge National Laboratory. We collected datasets at 250~mK and 550~mK with incident neutron wavelength $\lambda$~=~2.41~{\AA}~\cite{Xie2023stripe}.

\begin{figure}[tb]
    \center{\includegraphics[width=.99\linewidth]{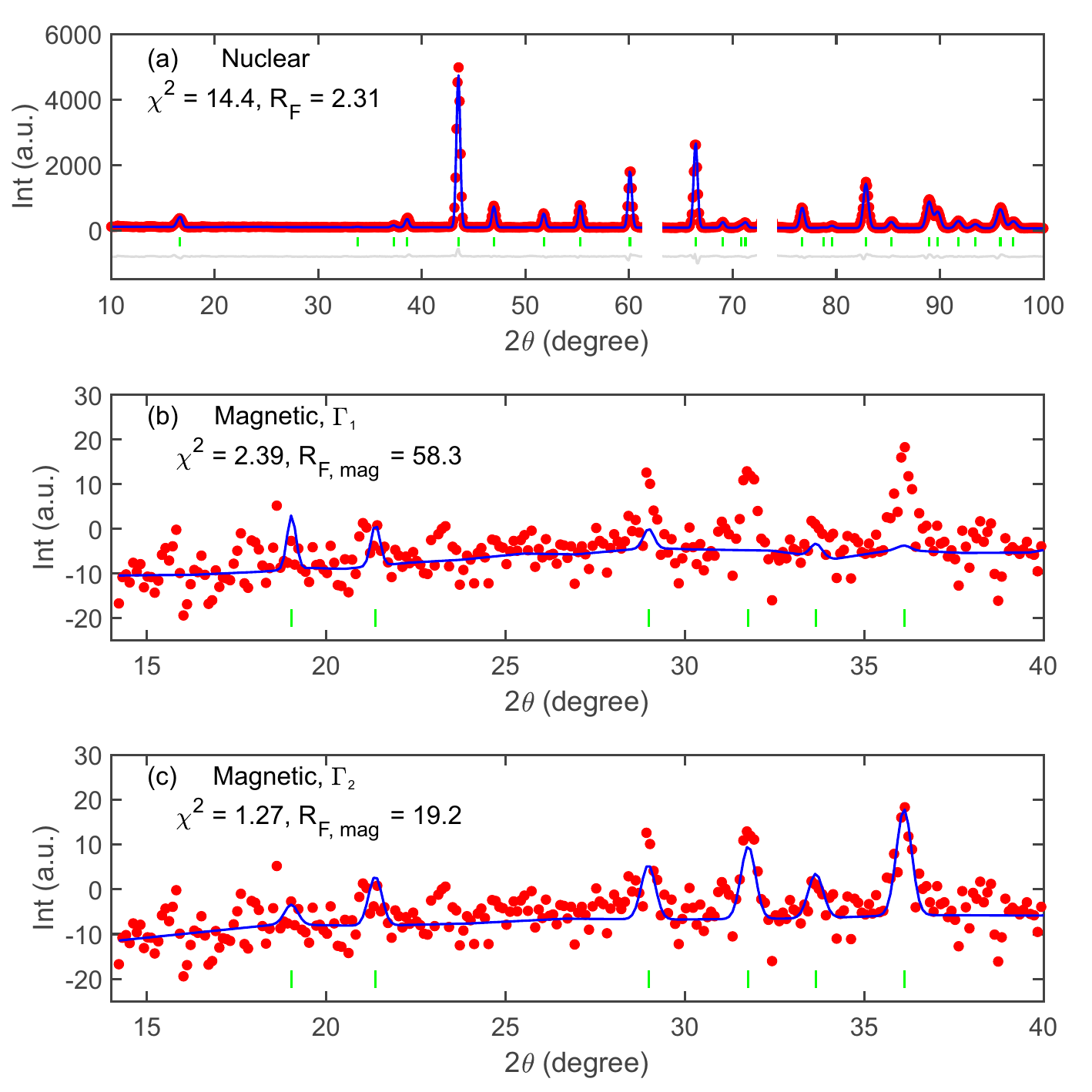}}
    \caption{~Neutron powder diffraction on \ccs.
    (a) Diffraction pattern collected at 0.55~K. Red dots show experimental data, blue and grey line represent the fitted curve and its difference with the experimental data. Green lines indicate position of the nuclear Bragg peaks. Narrow angular ranges around 62.2 and 73.3$^{\circ}$ are contaminated by contribution from Al and are masked out from the analysis.
    (b,c) Refinement of the difference profile, $I(T = 250$~mK$) - I(T = 550$~mK$)$, using $\Gamma_1$ (b) and $\Gamma_2$ irreps. Green lines indicate position of the magnetic Bragg peaks.
    }
    \label{Fig_SM_PND}
\end{figure}

As the first step we consider the dataset collected above $T_{\rm N}$. It can be nicely described using $R$-$3m$ space group \ccs\ and the lattice parameters $a = 4.40410(18)$~\AA, $c = 24.8546(11)$~\AA~[Fig.~\ref{Fig_SM_PND}(a)]. These results are consistent with previous reports~\cite{CCDC1952065, Xing20191}. By refining the nuclear scattering we determined all profile parameters that were fixed during the analysis of the magnetic structure. Under temperature decrease series of weak magnetic Bragg peaks arise at low scattering angles. They can be indexed using propagation wavevector $\mathbf{k}_{\rm mag} = (0, 1/2, 1)$. To visualize the magnetic scattering most clearly we made subtraction of 550 and 250~mK datasets [Figs.~\ref{Fig_SM_PND}(b,c)].

To determine the magnetic structure we used representation analysis approach. The representation analysis for $\mathbf{k}_{\rm mag}$ was performed using $\textsc{Mag2Pol}$~\cite{qureshi2019mag2pol} and $\textsc{SARAh}$~\cite{wills2000new} softwares. It yields two irreducible representations (ireps), $\Gamma = 1\Gamma_1 + 2\Gamma_2$, where $\Gamma_1$ and $\Gamma_2$ correspond to stripe-$x$ and stripe-$yz$ states. Basis vectors for $\Gamma_1$ is purely in-plane, $\psi = (4,0,0)$, while both, in- and out-of-plane components are allowed for $\Gamma_2$ [$\psi_1 = (2,4,0)$ and $\psi_2 = (0,0,4)$]. We tested our data against both models as shown in Fig.~\ref{Fig_SM_PND}(b,c). Clearly the experimental data can be describe much better using $\Gamma_2$ solution ($\chi^2 = 1.27$) as compared to $\Gamma_1$ ($\chi^2 = 2.39$). We note that the difference between the two solutions can be seen on the qualitative level, e.g. for the $\Gamma_1$ state magnetic reflections (0, 1/2, 1) [19.0$^{\circ}$] and (0, -1/2, 2) [21.4$^{\circ}$] should have stronger intensity than the peaks (1,-1/2,0) [29.0$^{\circ}$]  and (1,-1/2,3) [36.1$^{\circ}$], in a clear contradiction with the experimental data. On the other hand, $\Gamma_2$ solution accurately reproduces the observed pattern.
The fitted ordered moment lies primarily in the $ab$ plane, $m_{ab}$ = 0.67(9)~\mb\ with a statistically negligible out-of-plane canting, $m_{c}$ = 0.10(14)~\mb.
Complementary analysis performed using the single crystal neutron diffraction data confirms this conclusion. More details on the magnetic structure determination and its field-induced evolution can be found in Ref.~\cite{Xie2023stripe}.

\section{Details of ED}
\label{smsec::ed}

\begin{figure}[tb]
    \centering
    \includegraphics[width=0.45\textwidth]{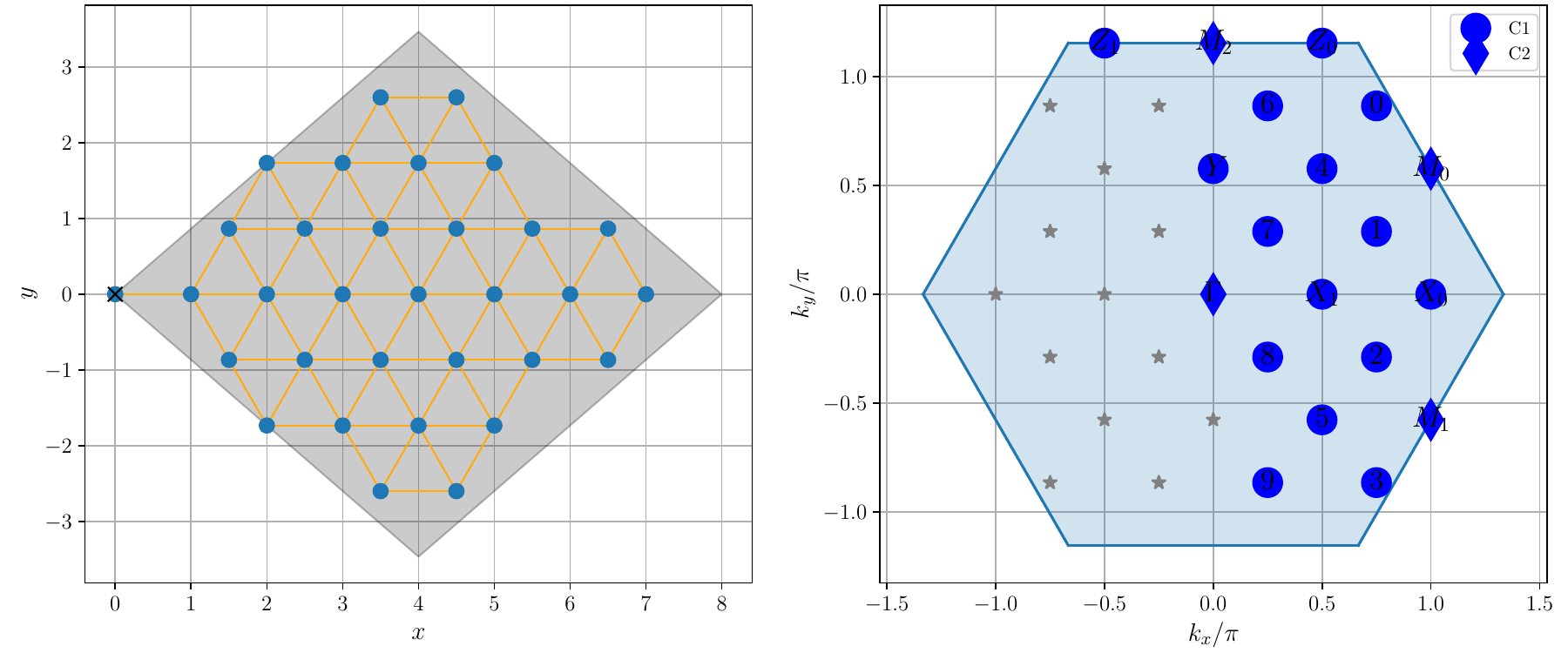}
    \caption{The $N=32$ lattice used in ED and the associated Brillouin zone.}
    \label{SM::fig::BZ}
\end{figure}

The Hamiltonian in Eq.~(\blue{1}) of the main text breaks all spin symmetries for non-trivial values of the couplings $J_{ij}$ and at finite magnetic fields. The only exploitable symmetry to be used in ED is the lattice translation symmetry. Figure~\ref{SM::fig::BZ} shows the finite-size lattice used throughout this paper. This $N=32$ sites cluster is particularly appealing for the three following reasons: i) it contains three discrete points on the experimentally accessible $\Gamma$ - M line, allowing for direct comparison between ED and INS data; ii) it contains all three M points, which allows us to represent all three stripe domains when the magnetic field vanishes; iii) the dimension of the Hilbert space in each sector, which is $\mathcal{O}(10^8)$, is sufficiently small to compute spectral functions at a reasonable cost.

With ED we compute the spectral functions $\mathcal{S}^{\alpha\beta}(\bold{Q}, \omega)$ which are the Fourier transforms in space and time of the spin-spin correlation functions,
\begin{equation}
\mathcal{S}^{\alpha\beta}(\bold{Q}, \omega) = \frac{1}{N} \sum_{i} e^{-\mathrm{i} \bold{Q}\cdot \bold{r}_i} \int_{-\infty}^{\infty} \mathrm{d}t \, e^{\mathrm{i}\omega t} \braket{S^{\alpha}_i(t) S^{\beta}_0(0)}.
\label{eq::Salphabeta_def_fourier_transform}
\end{equation}
In particular, because of the bond-dependent interaction, the off-diagonal spectral functions $\mathcal{S}^{\alpha\beta}, \ \alpha\neq\beta$ are non-vanishing.
The standard way of computing the spectral functions in Eq.~\eqref{eq::Salphabeta_def_fourier_transform} in ED is to obtain a sequence of poles and weights from the calculation of $\braket{A R(z) A^{\dagger}}$ with the resolvent $R(z) = (z - \mathcal{H} + E_0)^{-1}$ where $E_0$ is the ground-state energy and $A$ is the relevant spin operator (or combination thereof). These poles and weights are then broadened with a gaussian function having FWHM $\delta\omega = 0.04 \, \mathrm{meV}$ corresponding to the experimental resolution. We stress here that, for the off-diagonal spectral functions, the procedure requires a delicate matching of poles and weights with the diagonal terms~\cite{ohta_bogoliubov_1994,poilblanc_superconducting_2003}. To faithfully compare experimental data with ED results, we combine the spectral functions into the total dynamical structure factor (DSF) as
\begin{equation}
\mathcal{S}(\bold{Q}, \omega) = F(|\bold{Q}|) \sum_{\alpha\beta} g_{\alpha}g_{\beta} P_{\alpha\beta}(\bold{Q}) \mathcal{S}^{\alpha\beta}(\bold{Q}, \omega)
\label{SM::equ::S_q_omega}
\end{equation}
where $F(|\bold{Q}|)$ is the form factor associated to $\mathrm{Ce}^{3+}$ ions, $P_{\alpha\beta}(\bold{Q}) = \delta_{\alpha\beta} - Q_{\alpha}Q_{\beta}/\bold{Q}^2$ is the polarization factor and $g_{x} = g_{y} = g_{ab}, g_z$ are the Land\'e factors. Since $g_z \simeq 0.3 \ll g_{ab} = 1.77$, the diagonal $xx$ and $yy$ terms provide the major contribution as they are weighted by $g_{ab}^2$. Since the experimental results are along the $\Gamma$--Y--M line, we notice that, at the $\Gamma$ point, the polarization factor vanishes for the off-diagonal terms. Moreover, at the Y and M points at $B>0$, the only contributing off-diagonal terms are the $yz$ and $zy$ ones, which are weighted by $g_{ab} g_z$. After careful calculation, we observe that the off-diagonal contributions only account for about 6\% of the total DSF integrated over the magnetic field range $[0, 8] \ \mathrm{T}$ and energy window $[0, 1] \ \mathrm{meV}$ at the M point, and less at the Y point, and never bring any new feature which is observable in the experimental data. For the sake of simplicity we thus restrict to the diagonal terms only.

The last step for an accurate comparison of the experimental neutron scattering intensity with the ED data is to apply an overall rescaling factor on the DSF. We calibrate this factor based on the intensity of the magnon peak at the Y point of the BZ at 8~T.

\section{Fitting of exchange interactions}
\label{smsec::fit_couplings}

The minimal magnetic model of \ccs\ is described by Hamiltonian (\blue{1}) from the main text. This Hamiltonian contains five independent variables: $J, \, \Delta, \, J_{\pm\pm}, \, J_{z\pm}$ and $g_{ab}$. Moreover, from magnetic neutron diffraction we know that the magnetic ground state is the stripe-$yz$ state. This state can be stabilized not only by bond-dependent terms, but also by the next-nearest neighbor coupling, and therefore we also include $J_2$ in our analysis. Accurate quantitative determination of all exchange parameters is an ambiguous problem and we solve it by using LSWT in the high-field regime as well as at zero field. We then cross-check the obtained results by comparing the INS data with ED calculations for several representative sets of exchange parameters, which further allow us to verify the accuracy of our approach.

\begin{figure}[tb!]
\center{\includegraphics[width=1\linewidth]{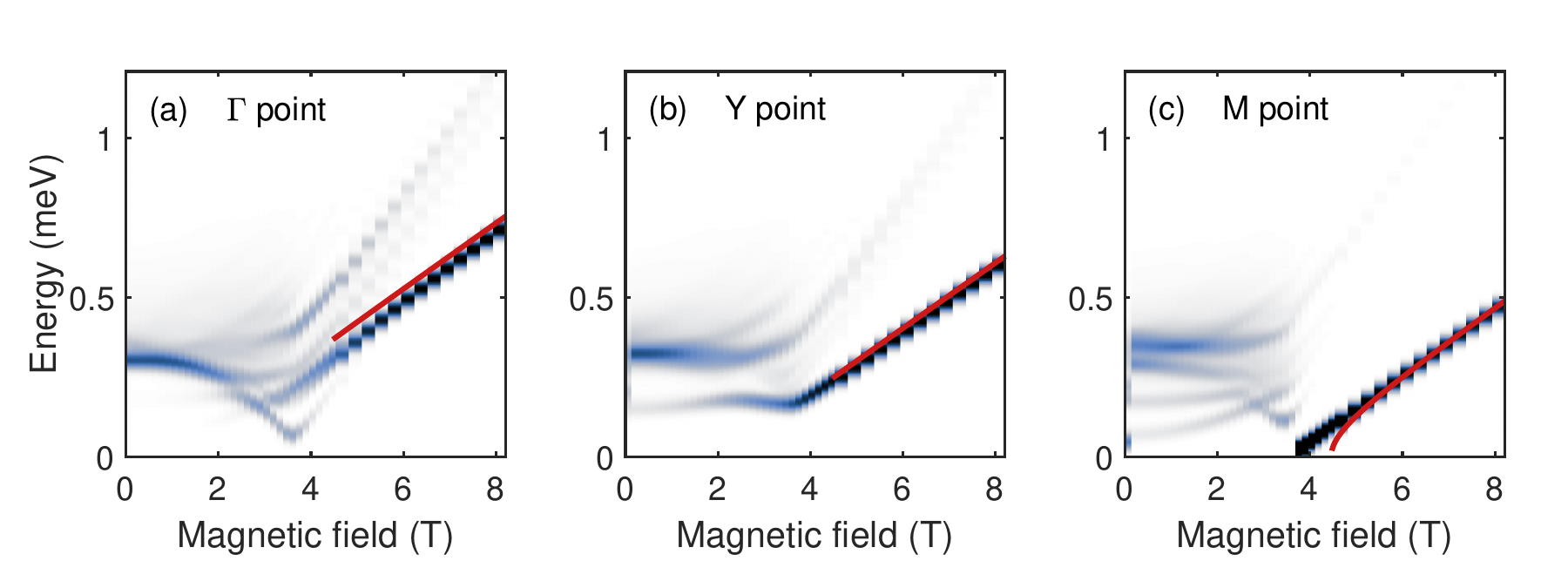}}
  \caption{~Dynamical structure factor of \ccs\ calculated by ED at multiple fields at (a)~$\Gamma$, (b)~Y and (c)~M points of the BZ. Red solid lines show calculations above the critical field using LSWT.
  }
  \label{Fig_SI_fit1}
  \vspace{-12pt}
\end{figure}

We start by determining the $g$-factor of the ground state doublet of Ce$^{3+}$ ion in \ccs. Usually, this can be done very accurately by ESR measurements or by measuring the field dependence of spin excitations in the field-polarized phase, where the magnetic mode should simply shift up in energy following the Zeeman law. However in close proximity to the phase transition between the stripe-$yz$ phase and the field-polarized phase, the linear dependence of the excitation energy versus magnetic field can be violated. To demonstrate this effect, in Fig.~\ref{Fig_SI_fit1} we plot the DSF calculated by ED and the magnon dispersions in the high-field regime from LSWT. One can see that at the Y point, LSWT and ED agree perfectly and the excitation follows the Zeeman law, while for the $\Gamma$ and M points the situation is more ambiguous. We therefore use our high-field INS data taken at Y to determine the $g$-factor and find $g_{ab} = 1.77$.

As the next step we proceed by determining the different components of exchange tensors using the high-field dispersion. It can be calculated analytically in LSWT as~\cite{li2018effect, avers2021finger}
\begin{align}
    \epsilon_{\bf k} = 3JS\sqrt{ A_{\bf k}^2 - |B_{\bf k}|^2}
    \label{Eq:LSWT_disp}
\end{align}
where $B_{\bf k} = B_{\bf k}' + iB_{\bf k}''$ and $A_{\bf k}$ and $B_{\bf k}$ are given by
\begin{align}
    A_{\bf k} &= h - 2(1+\alpha) + (1+{\Delta})(\gamma_{\bf k} + \alpha \gamma_{\bf k}^{(2)}) - 2\eta\gamma_{\bf k}' \\
    B_{\bf k}' &= -(1-{\Delta})(\gamma_{\bf k} + \alpha \gamma_{\bf k}^{(2)}) + 2\eta\gamma_{\bf k}'\\
    B_{\bf k}'' &= 2\mu\gamma_{\bf k}''
\end{align}
with $h = g_{ab}\mu_{\rm B} B/(3 J S)$, $\alpha = J_2/J$, $\eta = \Jpm/J$, $\mu = \Jzpm/J$ and the hopping amplitudes are given by
\begin{align}
    &\gamma_{\bf k} =
    \frac{1}{3} \sum_\beta \textrm{cos} \mathbf{k\delta}_{\beta} =
    \frac{1}{3}\bigg(\textrm{cos}k_x + 2 \textrm{cos}\frac{k_x}{2}\textrm{cos}\frac{\sqrt{3}k_y}{2}\bigg)  \\
    &\gamma_{\bf k}^{(2)} =
    \frac{1}{3} \sum_\beta \textrm{cos} \mathbf{k\delta}_{\beta}^{(2)} =
    \frac{1}{3}\bigg(\textrm{cos}\sqrt{3}k_y + 2 \textrm{cos}\frac{3k_x}{2}\textrm{cos}\frac{\sqrt{3}k_y}{2}\bigg)   \\
    &\gamma_{\bf k}' =
    \frac{1}{3} \sum_\beta \textrm{cos}\varphi_{\beta} \textrm{cos} \mathbf{k\delta}_{\beta} =
    \frac{1}{3}\bigg(\textrm{cos}k_x - \textrm{cos}\frac{k_x}{2}\textrm{cos}\frac{\sqrt{3}k_y}{2}\bigg)\\
    &\gamma_{\bf k}'' =  \frac{1}{3} \sum_\beta \textrm{sin}\varphi_{\beta} \textrm{cos} \mathbf{k\delta}_{\beta} =  \frac{1}{\sqrt{3}}\textrm{sin}\frac{k_x}{2} \textrm{sin}\frac{\sqrt{3}k_y}{2}
\end{align}
where $\delta_{\beta}$ are the bond vectors to nearest neighbors.

First, we constrain the next-nearest neighbor interaction parameter $J_2$. To do so we make use of the fact that for $k_x = 0$, $\gamma_{\bf k}, \gamma_{\bf k}'$ and $\gamma_{\bf k}''$ depend on $k_y$ as $\gamma \propto \textrm{cos} \frac{\sqrt{3}k_y}{2}$. In contrast, the term proportional to $J_2$, $\gamma_{\bf k}^{(2)}$, has a different functional form and therefore will change the shape of the dispersion away from a simple cosine. We found that at all relevant fields, $B \geq\ $ 5 T, the observed dispersion has negligible deviations from cosine behavior, which allows us to restrict $J_2/J \lesssim\ $0.03. Because of the small value of $J_2$ we fix it to zero in the following analysis. Note that the presence of substantial further nearest-neighbor terms, such as $J_3$ and $J_4$ can be also ruled out because of negligible deviations from cosine dispersion.

\begin{figure}[tb]
\center{\includegraphics[width=1\linewidth]{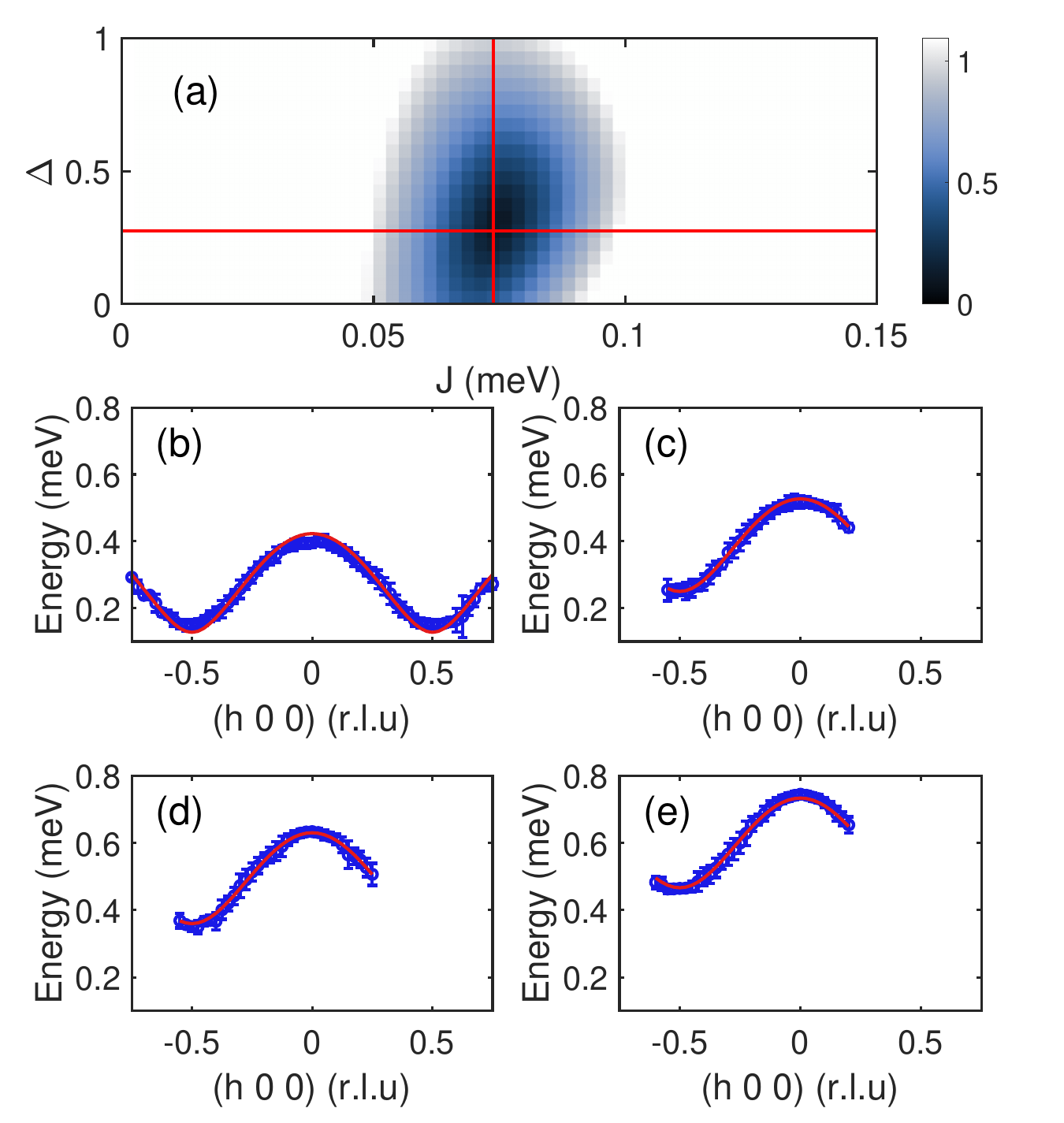}}
  \caption{~Fitting of $J$ and $\Delta$ for fixed $J_{\pm\pm}$ and $J_{z\pm}$.
  (a)~Fit quality, $\sqrt{ \sum_i (\frac{y_i - y_{\rm calc}}{e_i})^2 }$, in coordinates of $J$ and $\Delta$. Red cross indicates position of the global minimum.
  (b-e)~Magnon dispersion extracted from the INS data (dots) and results of the LSWT calculations for the best parameter set (solid lines) for different fields: 5~T (b), 6~T (c), 7~T (d) and 8~T (e).
  }
  \label{Fig_SI_fit2}
  \vspace{-12pt}
\end{figure}

We then fit the high-field dispersion using Eq.~\eqref{Eq:LSWT_disp}. We quantify the position of the magnon mode at multiple points in the reciprocal space for $B = 5, 6, 7, 8 \, \mathrm{T}$. However, such fitting does not allow us to equivocally identify a full set of couplings, but rather provides a mean of constraining two parameters out of the four remaining couplings. In our case, we use $J$ and $\Delta$ as dependent variables that are determined for every combination of $J_{\pm\pm}$ and $J_{z\pm}$ using least square fitting. As a representative example we show such fit results for the final parameter set in Fig.~\ref{Fig_SI_fit2}.

\begin{figure}
    \centering
    \includegraphics[width=.48\textwidth]{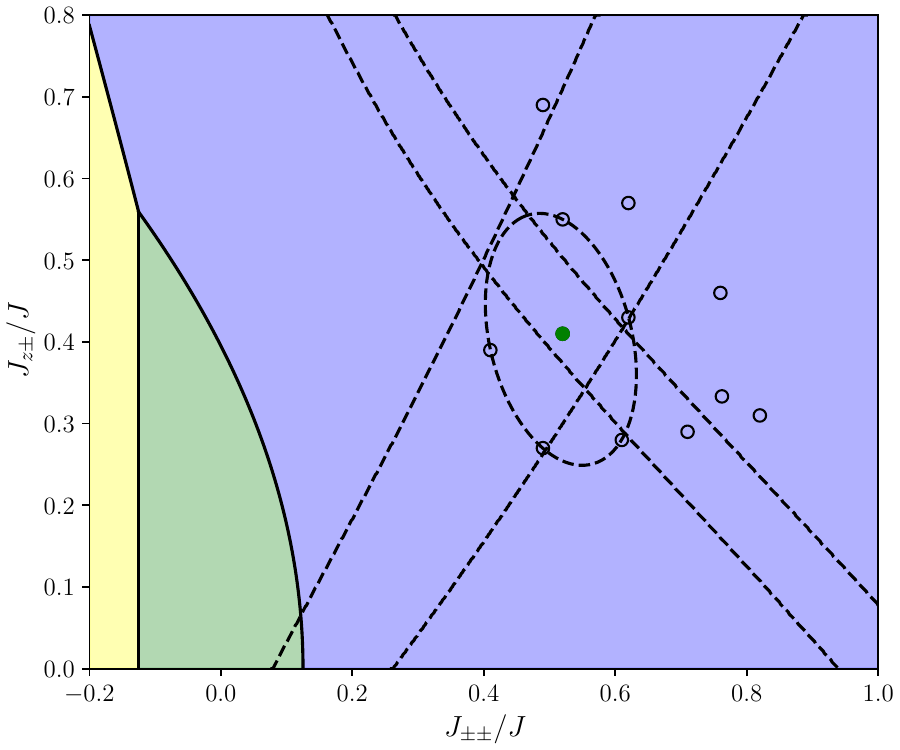}
    \caption{LSWT phase diagram for $\Delta=1/4$. The twelve couplings sets studied with ED are represented by small circles. The green circle is the chosen set, corresponding to the barycenter of the black dashed ellipse which encircles the phase space where \ccs\ could lie. Phase boundaries are from LSWT~\cite{maksimov2019anisotropic}.}
    \label{SM::fig::phase_diagram}
\end{figure}

The analysis so far leaves us with a series of coupling sets which, from the point of view of LSWT at high fields, provide a reasonable description of the features seen in the experimental data. To refine the couplings, we make use of the ED data at low field, since the comparison of experiment and LSWT at low field is more involved, as described in the main text. We thus compute the spectrum and the DSF in ED for twelve sets of couplings and focus the comparison between experiment and ED on the three following criteria: i) the value of the transition field between the stripe-$yz$ phase and the field-polarized phase; ii) the position (in energy) of the strong peak observed at $\Gamma$ at $0 \, \mathrm{T}$; iii) the discrepancy of the total DSF Eq.~\eqref{SM::equ::S_q_omega} at $\Gamma$, Y and M in the energy window $\left[0.2, 0.5\right] \, \mathrm{meV}$. This analysis allows us to remove six out of the twelve sets of couplings. With the current available data, we are not in a position to further narrow down the couplings. We thus decide to select the barycenter of the ellipsis in Fig.~\ref{SM::fig::phase_diagram} as the representative coupling set and obtain $J = 72.5 \, \mu\mathrm{eV}$, $\Delta = 0.25$, $\eta = 0.52$, $\mu = 0.41$, $g = 1.77$, $J_2 = 0$.

\begin{figure}
    \centering
    \includegraphics[width=.48\textwidth]{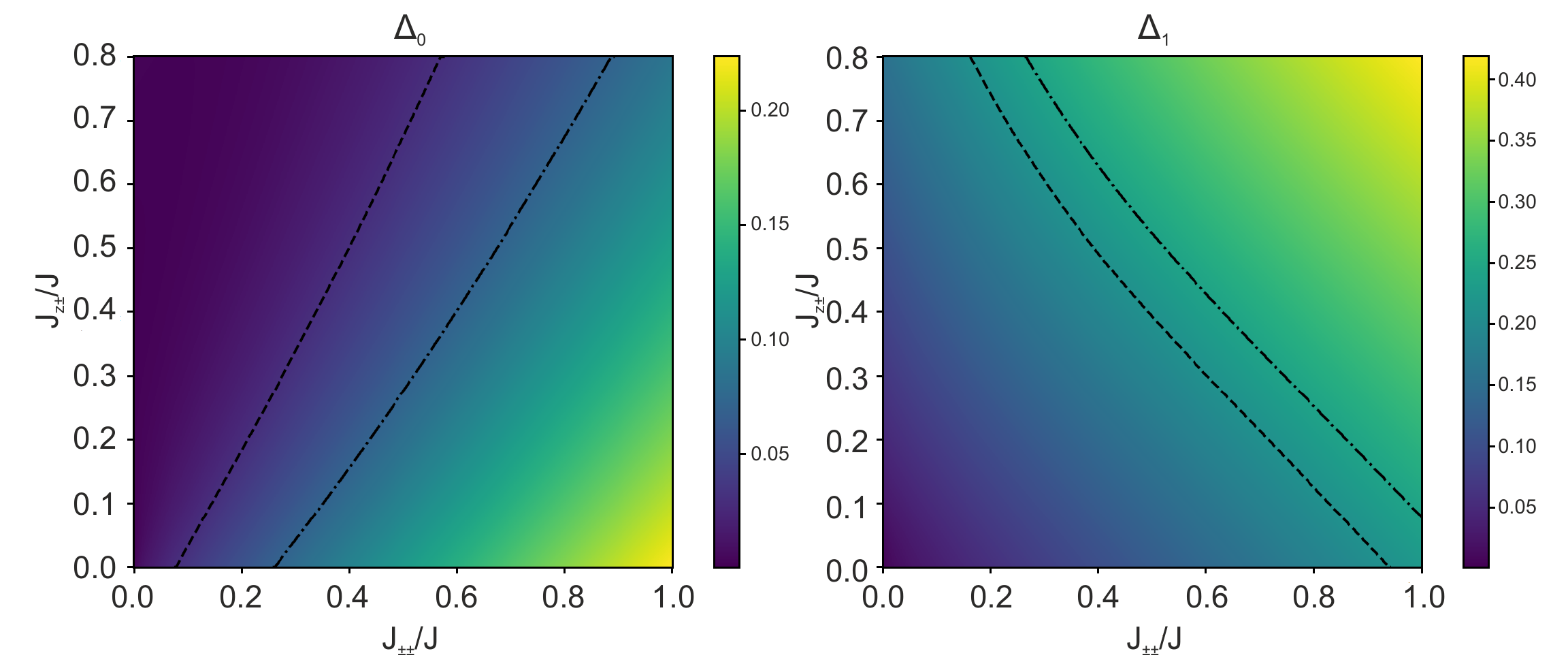}
    \caption{LSWT gaps at the M point of the BZ in vanishing magnetic field. The dashed lines are iso-gap values corresponding to (left) $\Delta_0/J = 0.69 \pm 0.14$ and (right) $\Delta_1/J = 3.12 \pm 0.07$ evaluated at $2\sigma$ and extracted from the INS data. These lines are further reported in the phase diagram Fig.~\ref{SM::fig::phase_diagram}.}
    \label{SM::fig::gaps}
\end{figure}

Finally, we cross check this analysis using LSWT at $B=0 \, \mathrm{T}$~\cite{maksimov2019anisotropic}. At the M point of the BZ the two magnon modes from the domains that are not present in finite field have a significant splitting. We can thus estimate lower and upper bounds for each of the associated gaps from the INS data and extract the iso-gap curves in $(J_{\pm\pm}, J_{z\pm})$ phase space for the corresponding LSWT gaps of the two modes. These are represented in Fig.~\ref{SM::fig::gaps} and further reported in Fig.~\ref{SM::fig::phase_diagram}, with the diamond-like crossing area corresponding to the viable phase space.

These phenomenological constraints on the two gaps at the M point are nearly orthogonal, yielding a rather tight region of the $J_{\pm\pm}$-$J_{z\pm}$ space, which includes the barycenter from the ED exploration discussed above, see Fig.~\ref{SM::fig::phase_diagram}. Thus, the two analyses are strongly supporting each other on the model parameter selection for {\ccs}.

\section{Linear spin wave theory at intermediate fields}
\label{smsec::lswt}

\begin{figure}
    \centering
    \includegraphics[width=.48\textwidth]{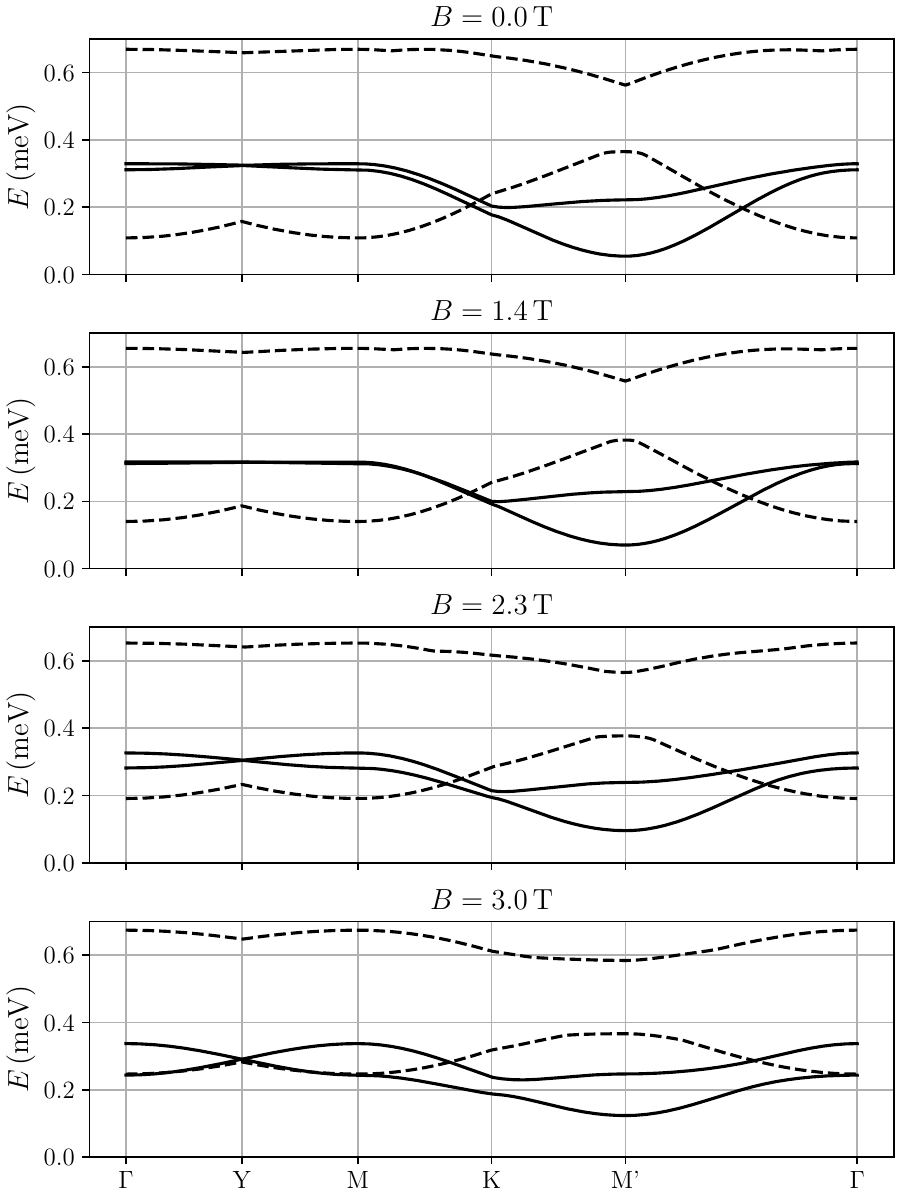}
    \caption{Linear spin-wave theory dispersion (full lines) of the magnon branches and two-magnon lower and upper thresholds (dashed lines) at low fields along the $\Gamma$ - Y - M - K - M' - $\Gamma$ path in the BZ, obtained using \textsc{SpinW}~\cite{spinw}.}
    \label{SM::fig::Energy_LSWT_path_BZ}
\end{figure}

At high fields and in a vanishing magnetic field, the LSWT dispersion is known analytically~\cite{maksimov2019anisotropic, avers2021finger}. At intermediate fields, however, the LSWT dispersion has, to our knowledge, not yet been analytically computed. We thus use \textsc{SpinW}~\cite{spinw} to extract the dispersion of the magnon modes and the lower and upper thresholds of the two-magnon continuum, Fig.~\ref{SM::fig::Energy_LSWT_path_BZ}.

\section{Stripe-$yz$ magnetic structure and out of plane tilt}
\label{smsec::stripeyz}
According to powder neutron diffraction data \ccs\ orders into stripe-$yz$ magnetic phase, see Sec.~\ref{smsec::NPD}. This phase is a collinear AFM state with moments pointing perpendicular to the nearest-neighbor bonds. In contrast to stripe-$x$, the stripe-$yz$ state allows of an out-of-plane canting of magnetic moments. In our model and is controlled by $J_{z\pm}$ term. In the LSWT approximation, it is given by~\cite{maksimov2019anisotropic}:
\begin{align}
    \mathrm{tan}(2\theta) = -4J_{z\pm}/[J(1-\Delta)+4J_{\pm\pm}].
    \label{eq::tilt}
\end{align}
Thus, in principle, the experimental value of the canting angle can provide another independent constrain to determine exchange Hamiltonian parameters. In our work we did not focus on this aspect for the following reasons:
From the analysis of the neutron diffraction data, we deduced $m_c =0.10(14) \mu_{\rm B}$ and the canting angle of the magnetic moment $12(16) \degree$~\cite{Xie2023stripe}. However, the statistical uncertainty is large and therefore we did not  include this parameter in our procedure of exchange parameters determination.

Our LSWT calculations that account for the out-of-plane canting clearly demonstrate that the spin canting does not affect topology of the magnetic excitations. We found that two magnon modes remain degenerate at the Y point at all fields up to saturation [see green line in Fig.4 (b) and black lines in Figs.2 (b-f)], regardless of the presence of such a spin canting.

\section{Non-linear SWT}
\label{Sec_NLSWT}

Here we provide some details of the non-linear SWT leading to the results  in Fig.~3(a) of the main text.

\subsection{General formalism}
\vskip -0.2cm

Within the $1/S$  spin-wave expansion, the  reference frame on each site
is rotated to a local one  with the  $z$ axis pointing along the spin's quantization axis
given by the classical energy minimization, with $\widetilde{\mathbf{S}}_i$ being the spin vector in the local reference frame at the site $i$ and generic nearest-neighbor spin Hamiltonian given by
\begin{align}
\mathcal{H} = \sum_{\langle ij\rangle} \widetilde{\mathbf{S}}_i^{\rm T}
\widetilde{\bf J}_{ij} \widetilde{\mathbf{S}}_j\, ,
\label{eq_Hij_loc}
\end{align}
where the ``rotated'' exchange matrix is
\begin{align}
\widetilde{\bf J}_{ij} = \left(
\begin{array}{ccc}
\widetilde{J}_{ij}^{xx} &\widetilde{J}_{ij}^{xy} &\widetilde{J}_{ij}^{xz} \\
\widetilde{J}_{ij}^{yx} &\widetilde{J}_{ij}^{yy} &\widetilde{J}_{ij}^{yz} \\
\widetilde{J}_{ij}^{zx} &\widetilde{J}_{ij}^{zy} &\widetilde{J}_{ij}^{zz}
\end{array} \right).
\label{eq_Jij_rot}
\end{align}
The LSWT  needs only diagonal and $\widetilde{J}_{ij}^{xy}(\widetilde{J}_{ij}^{yx})$ terms.

The non-linear anharmonicities are of two types. The so-called quartic one comes from the same terms as the LSWT in the higher $1/S$-order of expansion of the bosonic representation for spin operators.  As is argued below, it is of the lesser importance for our problem and will be ignored. The second type of anharmonicity is  the three-magnon interaction from the two off-diagonal terms in (\ref{eq_Jij_rot}) that couple $z$ with $x(y)$ spin components
\begin{align}
\label{H3}
\mathcal{H}_{\rm od}=\sum_{\langle ij \rangle}\left( \widetilde{J}_{ij}^{xz}
\widetilde{S}^x_i \widetilde{S}^z_j +\widetilde{J}_{ij}^{yz}
\widetilde{S}^y_i \widetilde{S}^z_j +\{i\leftrightarrow j\}\right).
\end{align}
The Holstein-Primakoff bosonization of Eq.~\eqref{H3} yields the three-particle interaction
\vskip -0.05cm
\noindent
\begin{align}
\label{H3HP}
\mathcal{H}_3= \sum_{\langle ij \rangle}
\left( \widetilde{V}_{ij}  a^\dagger_i a^\dagger_j a^{\phantom \dagger}_j+\text{H.c}+\{i \rightarrow j\}\right),
\end{align}
\vskip -0.1cm
\noindent
where $\widetilde{V}^{}_{ij}\!=\! -\sqrt{S/2} \left( \widetilde{J}_{ij}^{xz}+\mathrm{i}\widetilde{J}_{ij}^{yz}\right)$.

The  general procedure of obtaining   symmetrized three-magnon interactions
for the quasiparticles of the normal modes of the corresponding ground state from the non-diagonal Holstein-Primakoff form of Eq.~(\ref{H3HP}) typically requires numerical diagonalization of the harmonic LSWT Hamiltonian and is also technically quite involved otherwise, see Ref.~\cite{smit2020magnon}. However
the resultant form of the important decay part of it is quite general,
\begin{eqnarray}
\mathcal{H}_3\!=\!\frac{1}{2\sqrt{N}}\!\sum_{{\bf k}+{\bf q}=-{\bf p}}\sum_{\eta\nu\mu} \left(
\widetilde{V}^{\eta\nu\mu}_{{\bf q}{\bf k};{\bf p}}
d^{\dagger}_{\eta{\bf q}} d^{\dagger}_{\nu{\bf k}} d^{\phantom{\dag}}_{\mu-{\bf p}}+{\rm H.c.}\right)\!, \ \ \ \
\label{Hdecay}
\end{eqnarray}
where $d^{\dagger} (d)$ are magnon operators of the true quasiparticles, indices $\eta$, $\nu$, and $\mu$ numerate their branches, and $\widetilde{V}^{\eta\nu\mu}_{{\bf q}{\bf k};{\bf p}}$ is the symmetrized vertex.
With this interaction (\ref{Hdecay}),  standard diagrammatic rules allow for a systematic calculation
of the quantum corrections to the magnon spectra in the form of  self-energies $\Sigma^\mu (\mathbf{k},\omega)$
\begin{eqnarray}
\Sigma^{\mu}({\bf k}, \omega) = \frac{1}{2N}\sum_{{\bf q},\nu\eta}
\frac{|\widetilde{V}^{\eta\nu\mu}_{{\bf q}{\bf k};{\bf p}}|^2}
{\omega - \varepsilon_{\nu,\bf q} - \varepsilon_{\eta,{\bf k}-{\bf q}}+\mathrm{i}\delta}\,,
\label{Sigma}
\end{eqnarray}
where $\varepsilon_{\nu,\bf q}$ are magnon energies.

Ignoring contributions from the aforementioned quartic  and the so-called source cubic term that do not contribute to decay processes, the magnon Green's function for the branch $\mu$ can be approximated as
\begin{eqnarray}
G^{-1}_\mu({\bf k},\omega)\approx\omega-\varepsilon_{\mu,{\bf k}} -
\Sigma^{\mu}({\bf k}, \omega)\,,
\label{GF}
\end{eqnarray}
which allows us to evaluate the corresponding spectral function
$A_\mu({\bf k},\omega)=-(1/\pi){\rm Im}G_\mu({\bf k},\omega)$.

\subsection{Non-linear SWT for our model}
\vskip -0.2cm

\begin{figure*}[t]
\includegraphics[width=0.49\linewidth]{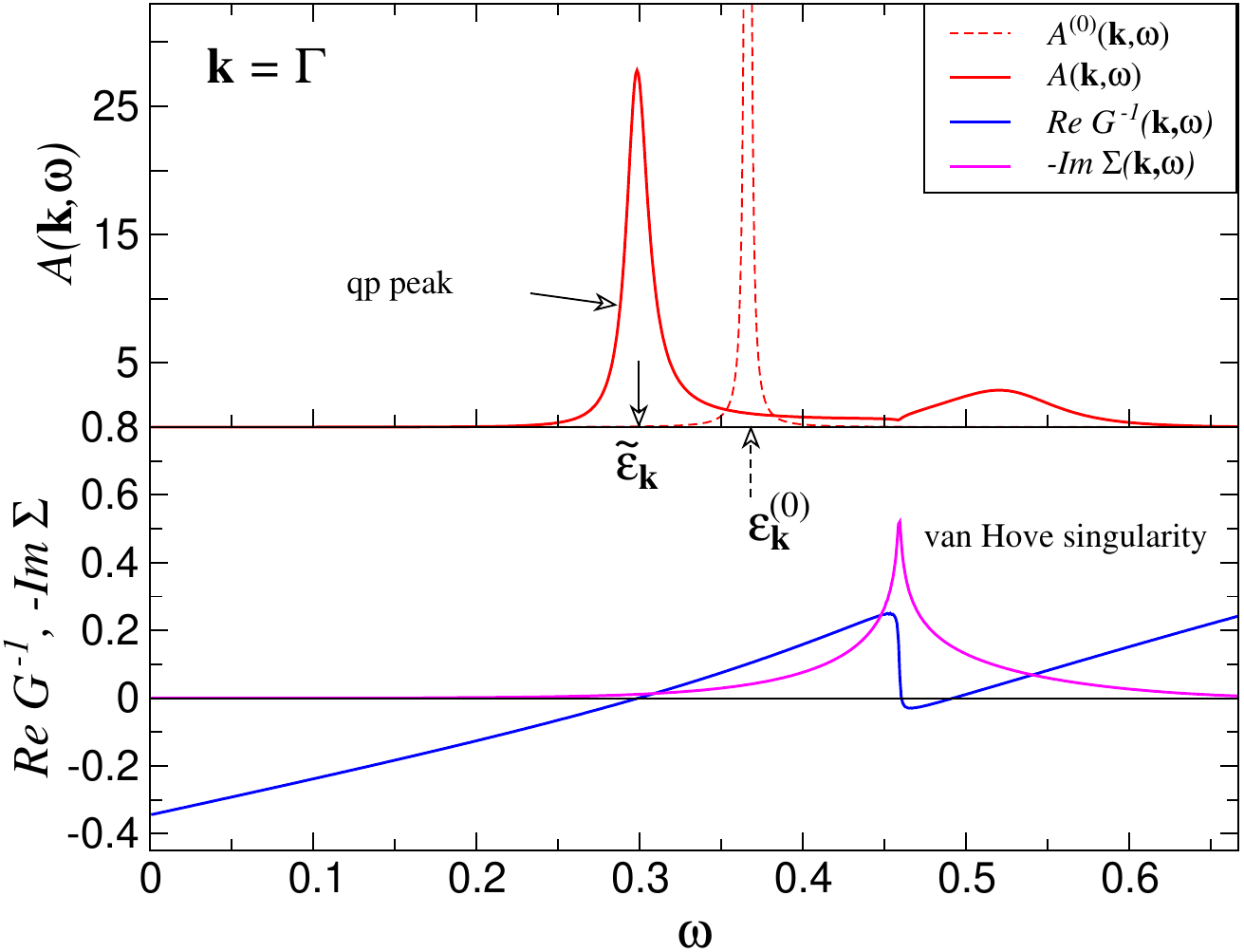} \ \
\includegraphics[width=0.49\linewidth]{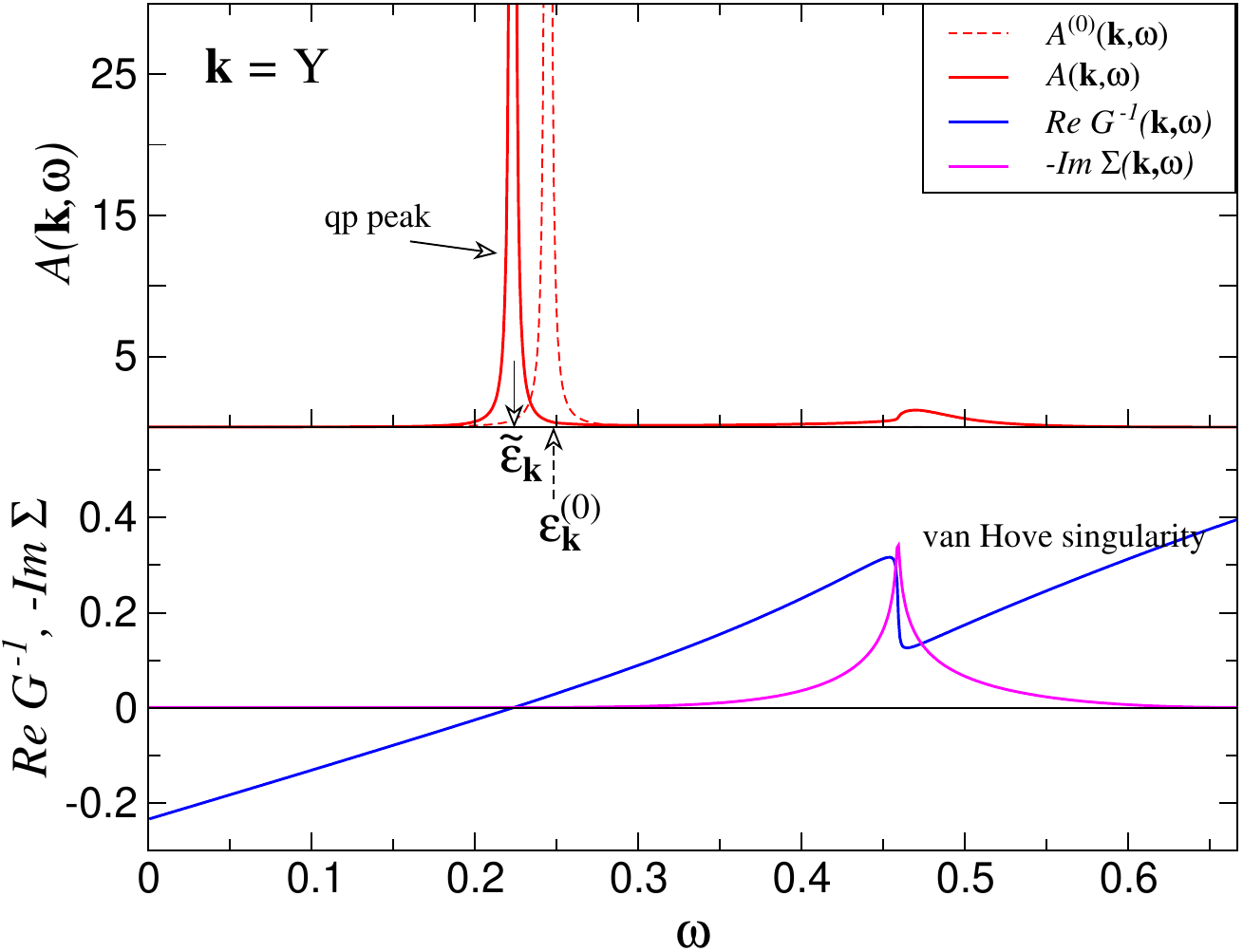}
\vskip -0.2cm
\caption{Magnon spectral functions $A({\bf k},\omega)$ at the (left) $\Gamma$ and (right) Y points, for $B\!=\!B_c$,
upper panels; Real and imaginary parts of $G^{-1}({\bf k},\omega)$, lower panels.
Dashed and solid arrows indicate positions of the LSWT and renormalized magnon peaks,  respectively.}
\label{fig_1DAwk}
\vskip -0.4cm
\end{figure*}

{\it Quartic terms ---} The quartic terms give the $1/S$-corrections to the spectrum  via the Hartree-Fock decoupling of them down to the harmonic terms, see, e.g., Ref.~\cite{Chernyshev2009}. Therefore, one can use  $\delta S$,   the deviation of the ordered moment from the saturated value $S\!=\!1/2$, which is given by one of such Hartree-Fock averages, as an estimate of the contribution of the quartic terms.

In zero field, throughout the stripe-$yz$ phase of the model~(1) of the main text, one finds a highly ordered state with $\langle S\rangle\!\agt\!0.45$ for the relevant range of parameters~\cite{maksimov2019anisotropic}. We can also calculate the ordered moment for $B\!>\!B_c$ for the parameters discussed  above and show that it reaches the minimum value of $\langle S\rangle\!=\!0.457$ at $B_c$, giving maximal $\delta S/\langle S\rangle\!\alt\!10\%$. Thus, one can estimate the contribution of the quartic terms to the spectrum renormalization as being of the same order and neglect it.

{\it Cubic terms for $B\!\geq\!B_c$ ---} We can make progress with the effect of the anharmonic terms in the spectrum of \ccs\ at the critical field $B\!=\!B_c$ and above. In this case, the nonlinear SWT formalism described in the previous section simplifies considerably as the magnon excitation spectrum consists of only one mode.

The LSWT diagonalization yields the magnon energy given in Sec.~\ref{smsec::fit_couplings}, while the decay term simplifies to,
\begin{eqnarray}
\mathcal{H}_3\!=\!\frac{1}{2\sqrt{N}}\!\sum_{{\bf k},{\bf q}} \left(
\widetilde{V}^{(3)}_{{\bf k},{\bf q}}
a^{\dagger}_{{\bf q}} a^{\dagger}_{{\bf k}-{\bf q}} a^{\phantom{\dag}}_{{\bf k}}+{\rm H.c.}\right)\!,
\label{HdecayHsat}
\end{eqnarray}
with the vertex given by,
\begin{eqnarray}
\widetilde{V}^{(3)}_{{\bf k},{\bf q}} = -\sqrt{6S}\left( 2J_{\pm\pm}+\mathrm{i} J_{z\pm}\right)\Phi_{{\bf k},{\bf q}}\,,
\label{V3}
\end{eqnarray}
where the kinematic part is
\begin{eqnarray}
\Phi_{{\bf k},{\bf q}}=
\sin\frac{q_x}{2}\sin\tilde{q}_y+\sin\frac{k_x-q_x}{2}\sin(\tilde{k}_y-\tilde{q}_y)\,,
\label{Phi3}
\end{eqnarray}
with the shorthand notation $\tilde{p}_y\!=\!\sqrt{3}p_y/2$.

{\it Spectral function ---} With these, using the self-energy in~\eqref{Sigma} and the approximate magnon Green's function in~\eqref{GF}, calculation of the magnon spectral function in the $({\bf k},\omega)$-plane presented in Fig.~3(a) of the main text is straightforward. To substantiate these results, in Fig.~\ref{fig_1DAwk} we show the 1D $\omega$-cuts for such a spectral function for the two representative points, $\Gamma$ and Y, for $B\!=\!B_c$. There is a clear downward renormalization of the magnon mode at the  $\Gamma$ point, which becomes much smaller  at the Y point, in close agreement with both ED and experimental data. The origin of this effect is a repulsion from the saddle-point-like Van Hove singularity in the two-magnon continuum, indicated in the lower panels in Fig.~\ref{fig_1DAwk}.

There is a substantial broadening of the quasiparticle peak in the $\Gamma$ point spectrum, but the main effect is the downward shift. There is also a transfer of the spectral weight into the higher-energy tail of the spectrum due to the one-to-two-magnon coupling, contributing to the ``extra mode'' at higher energy in the same ${\bf k}$-region where the magnon is most suppressed. The impact of the coupling between the single- and two-magnon sectors is diminishing near the Y point because of the kinematic structure of the vertex~\eqref{V3} and vanishing density of states of the two-magnon continuum at its boundary.

{\it Further analysis ---} Two additional comments are in order. First, the cubic anharmonicity~\eqref{V3} for $B\!\geq\!B_c$  is solely due to the bond-dependent terms, because of the collinear spin directions in the field-polarized state. The same is true for the zero-field stripe-$yz$ phase. While the  nonlinear SWT formalism is significantly more complicated in the latter case because of the two-sublatice order in the stripe-$yz$ phase, one can compare the magnitude of the decay vertex~\eqref{Hdecay} in it with the vertex in \eqref{V3}. One key difference is that, in the field-polarized state, the two bond-dependent terms simply add up with the same kinematic part~\eqref{Phi3}. This is not so for the zero-field vertex, resulting in about ten-fold smaller overall probability $\big|\widetilde{V}^{(3)}_{{\bf k},{\bf q}}\big|^2$ that  contributes to the self-energy in~\eqref{Sigma}.

Absent of the explicit calculation of the zero-field spectrum renormalization, this analysis is generally in agreement with the observation of only minor deviations of the zero-field spectrum from the LSWT results. It also seems to explain why  parameter searches described in Sec.~\ref{smsec::fit_couplings} yielded  very close target regions of the parameter space using the quasiclassical LSWT analysis and quantum ED approach.

Second, the evolution of the nonlinear corrections to the spectrum from the minor ones in zero field to substantial at $B\!=\!B_c$ is not just interpolation and is likely more involved and nuanced. The reason is the field-induced spin non-collinearity  in the canted stripe-$yz$ phase for  $B\!\leq\!B_c$, which necessarily involves contributions from the exchange terms to the anharmonic couplings in~\eqref{H3}. These are known to induce unexpectedly strong modifications in the spectrum in the otherwise nearly classical states even in much simpler models; see Refs.~\cite{luscher2009exact,zhitomirsky1999instability}. Again, this  observation is in a general agreement with the discussion of the experimental data and results by ED that demonstrate such strong quantum effects in the spectrum of \ccs\ for  $B\!\leq\!B_c$.\\